\documentclass[12pt,preprint]{aastex}
\usepackage{natbib,amsmath}

\catcode`\@=11
\def\gsim{\ifmmode{\mathrel{\mathpalette\@versim>}}
    \else{$\mathrel{\mathpalette\@versim>}$}\fi}
\def\lsim{\ifmmode{\mathrel{\mathpalette\@versim<}}
    \else{$\mathrel{\mathpalette\@versim<}$}\fi}
\def\@versim#1#2{\lower 2.9truept \vbox{\baselineskip 0pt \lineskip
    0.5truept \ialign{$\m@th#1\hfil##\hfil$\crcr#2\crcr\sim\crcr}}}
\catcode`\@=12

\newcommand{\chandra}{{\it Chandra\/}}

\begin{document}

\slugcomment{Submitted to ApJ}

\title{AGN activity and the misaligned hot ISM in the compact  radio elliptical NGC4278}

\author{S. Pellegrini$^{1}$, J. Wang$^2$, G. Fabbiano$^2$, D.W. Kim$^2$,
N.J. Brassington$^3$, J.S. Gallagher$^4$, 
G. Trinchieri$^5$, A. Zezas$^6$}
%\footnote{E-mail:  silvia.pellegrini@unibo.it}

\affil{$^1$Department of Astronomy, University of Bologna, 
                       via Ranzani 1, 40127 Bologna, Italy \\
$^2$Harvard-Smithsonian Center for Astrophysics, 60 Garden St, 
Cambridge, MA 02138, USA \\
$^3$School of Physics, Astronomy and Mathematics, University of Hertfordshire, Hatfield, UK\\
$^4$Department of Astronomy, University of Wisconsin-Madison, Madison, WI 53706, USA \\
$^5$INAF-Osservatorio Astronomico di Brera, Via Brera 28, 20121 Milano, Italy\\
$^6$Physics Department, University of Crete, Heraklion, Greece  }

\begin{abstract} 
The analysis of a deep (579 ks) $Chandra$
  ACIS pointing of the elliptical galaxy NGC4278, which hosts a low
  luminosity AGN and compact radio emission,  allowed
  us to detect extended emission from hot gas out to a
  radius of $\sim 5$ kpc, with 0.5--8 keV luminosity of $2.4\times
  10^{39}$ erg s$^{-1}$. The emission is elongated in the NE-SW direction, 
  misaligned with respect to the stellar body, and aligned with the ionized gas, and with the $Spitzer$
  IRAC 8$\mu m$ non-stellar emission.  The nuclear X-ray luminosity
  decreased by a factor of $\sim 18$ since the first $Chandra$
  observation in 2005, a dimming that enabled the detection of hot gas
  even at the position of the nucleus. Both in the projected and
  deprojected profiles, the gas shows a significantly larger
  temperature ($kT=0.75$ keV) in the inner $\sim 300$ pc than in
  the surrounding region, where it stays at $\sim 0.3$ keV, a value
  lower than expected from standard gas heating assumptions.  The
  nuclear X-ray emission is consistent with that of a low radiative
  efficiency accretion flow, accreting mass at a rate close to the
  Bondi one; estimates of the power of the nuclear jets require that
  the accretion rate is not largely reduced with respect to the Bondi
  rate.  Among possibile origins for the central large hot gas
  temperature, such as gravitational heating from the central massive black
  hole and a recent AGN outburst, the interaction with the nuclear
  jets seems more likely, especially if the latter  remain
  confined, and heat the nuclear region frequently.  The unusual
  hot gas distribution on the galactic scale could be due to 
  the accreting cold gas triggering the cooling of the hot phase, a
  process also contributing to the observed line emission from ionized
  gas, and to the hot gas
 temperature being   lower than expected; alternatively, the latter
 could be due to an efficiency of the type Ia supernova energy mixing 
lower than usually adopted.
\end{abstract}


\section{Introduction}

In recent years, high angular resolution X-ray observations of
early type galaxies performed with $Chandra$ 
allowed us to investigate extensively the connection between the central
supermassive black hole (MBH) and the surrounding hot interstellar
medium (ISM; e.g., Forman et al. 2007, Allen et al. 2006, Diehl \&
Statler 2008, David et al. 2009, Million et al. 2010), a study with
important implications for the building of a complete
picture of the host galaxy--MBH coevolution (Silk \& Rees 1998, Di
Matteo, Springel \& Hernquist 2005, Hopkins et al. 2005, Merloni \&
Heinz 2008, Cattaneo et al. 2009, Schawinski et al. 2009, Johansson et
al. 2009, Ciotti et al. 2010, Kaviraj et al. 2011).  The
investigations so far concentrated mostly on X-ray bright and massive
ellipticals, where the high hot gas surface brightness most clearly
reveals signs of feedback heating as cavities, shells, filaments, and
edges.  For a complete understanding of the MBH-host galaxy
coevolution process, it is however important to realize how the
activity cycle works even in low to intermediate mass galaxies, which are
far more numerous, and currently almost unexplored.  Other
aspects still unclear include the relationship between the nuclear
emission and other galactic properties, as the fuel availability for
the MBH (e.g., Ho 2008, Pellegrini 2010, Gallo et al. 2010); the
accretion modalities [standard disk plus hot corona, Haardt \&
  Maraschi (1993); or radiatively inefficient accretion, RIAF, Narayan
  \& Yi (1995); e.g., Ptak et al. 2004, Maoz 2007, Ho 2008]; and what determines the
partition in the accretion output between radiative and mechanical
energies (e.g., Allen et al. 2006, Merloni \& Heinz 2007).

In this paper we analyze two new $Chandra$ pointings of the medium-mass
elliptical galaxy NGC4278 ($d=16.1$ Mpc\footnote{At this distance
  $1^{\prime\prime}=78$ pc.}, and $L_B=1.7\times
10^{10}L_{B,\odot}$; Tab.~\ref{tab1}), obtained in March 2010; we also
consider the data of 6 previous pointings 
between 2005 and 2007, totalizing 579 ks of net exposure time.  With this deep
observation we explore the origin of the nuclear emission and the
possible interaction of the AGN activity with the surrounding hot ISM.
The spatially resolved properties of the low surface brightness hot
gas could be determined at a high level of accuracy thanks to the long
exposure time, the careful subtraction of the emission from stellar
sources allowed for by the $Chandra$ high angular resolution, and the
significant dimming of the nucleus during year 2010;
all this has allowed us to determine the gas properties of the central region 
(down to within a radius of $2^{\prime\prime}$ from the nucleus) and
of the more extended galactic body, out to a
radius of $\sim 70^{\prime\prime}$ ($\sim 5$ kpc).

NGC4278 is a relatively isolated elliptical galaxy, belonging to the
LGG 279 group of 17 galaxies (Garcia 1993); its morphological type is
E1-2, it has a regular optical shape, and a uniformely old stellar
population (Shapiro et al. 2010, Kuntschner et al. 2010).  NGC4278
also hosts a LINER1 nucleus, with a weak broad component in the
optical $H\alpha$ line (Ho et al. 1997).  Unlike standard AGNs,
NGC4278 shows an X-ray nuclear emission of low level ($\sim 10^{40}$
erg s$^{-1}$; Younes et al. 2010), and a very sub-Eddington bolometric
emission ($5\times 10^{-6}$, Eracleous et al. 2010a, Nemmen et
al. 2011).  This LINER is known to be radio-loud (Terashima \& Wilson
2003), and has been deeply studied in the radio.  At arcsecond
resolution, the VLA shows an unresolved emission from 5 to 43 GHz
(Wrobel \& Heeschen 1984, Nagar et al. 2001), while an extended
structure is visible at milliarcsecond resolution with the VLBI at 18
and 6 cm (Schilizzi et al. 1983), and a parsec-scale two-sided radio
jet was resolved by the VLBA emerging from a central compact component
(Giroletti et al. 2005).  The highly bent jet morphology suggests that
the jet may be interacting with the ambient medium in the inner few
parsecs.  The total radio (VLA) luminosity is $P_{1.4GHz}=1.0\times 10^{21.7}$ W
Hz$^{-1}$ (Giroletti et al. 2005, for our adopted distance), and then 
at least 2 orders of magnitude lower than in
powerful radio-loud AGN.

NGC4278 finally hosts a multiphase interstellar medium. The SAURON
survey (Sarzi et al. 2006) revealed strong emission from ionized gas
in the central region, whose kinematical structure is consistent with
that inferred for a massive ($M_{HI}=6.9\times 10^8$M$_{\odot}$),
extended (with a diameter of 37 kpc), and regular HI disk (Raimond et
al. 1981, Morganti et al. 2006).  CO emission was not detected and an
upper limit to the molecular gas mass of $M_{H_2}<6.9\times
10^6$M$_{\odot}$ was placed (Crocker et al. 2011). Recently, Tang et
al. (2011) presented mid-infrared emission maps of the extended
ionized gas, the warm molecular hydrogen and the dust, using $Spitzer$
spectroscopic observations.

In this paper we present the deep \chandra\ observations and their
imaging and spectral analysis in Sects. 2, 3 and 4; we summarize the
observational results in Sect. 5, discussing next the origin of the
nuclear emission (Sect. 5.1), that of the hot gas discovered at the
nucleus (Sect. 5.2), and the relationship between the various gas
phases (Sect. 5.3). We finally present our conclusions in Sect. 6.

\section{Observations and data analysis}

We utilized 6 archival and 2 new \chandra\ observations of NGC 4278,
imaged with the S3 chip of \chandra\ Advanced CCD Imaging Spectrometer
(ACIS; Garmire et al. 2003).  The two recent observations (PI:
Fabbiano) were taken on March 15, 2010 (obsID 11269) and March 20,
2010 (obsID 12124), with exposure times of 81.9 ks and 25.8 ks,
respectively.

We reduced and analyzed the observational data using the
\chandra\ Interactive Analysis of Observations (CIAO)
tools\footnote{See http://cxc.harvard.edu/ciao/ for details on
  CIAO.}. The  CIAO script {\sc chandra\_repro} was used to reprocess all
data with the latest calibration, remove pixel randomization, and
apply the sub-pixel (``EDSER'') algorithm. We reprojected the event
lists to obsID 11269 and merged the observations to create a master
event file and the associated exposure map using {\sc merge\_all}.
After a thorough source searching with {\sc wavdetect} (Freeman et
al. 2002) using all 8 co-added observations, 251 X-ray point sources
were found.  Details of the procedures performed to detect point
sources were reported in Brassington et al. (2009) and Fabbiano et
al. (2010).

From the co-added data set, \chandra\ images in 0.3-0.9 keV (``soft
band''), 0.9-2.5 keV (``medium band''), and 2.5-8.0 keV (``hard
band'') were created and adaptively smoothed using the CIAO task
$csmooth$; the smoothing kernel was constrained to yield a minimum
signal-to-noise ratio (S/N) of 2.5$\sigma$ and a maximum of 5$\sigma$.  An
X-ray ``false color'' image was created by combining these three
smoothed and exposure corrected images shown in red, green, and blue,
respectively (Figs.~\ref{f1}a and 1b).

To study the extended emission, we first masked out all point sources
using apertures with $2\arcsec$ radii, corresponding to $\approx$95\%
PSF encircled-energy radius at 1.496 keV.  The resulting ``swiss
cheese'' image should include a minimum contribution from detected point
sources.  The spectral extraction was done with the CIAO script
$specextract$\footnote{See CIAO thread
  \url{http://cxc.harvard.edu/ciao/threads/specextract/}.}, which
creates area-weighted Response Matrix Files (RMF) and Ancillary
Response Files (ARF) for each region in individual observations.
Background counts were taken from an annulus of source-free region
between $r=100\arcsec$ and $r=120\arcsec$ (see Sect. 3.1 on
determining the extent of the diffuse emission), and within the ACIS-S3
chip in all observations.  For each region of interest, source spectra
were grouped to a minimum of 15 counts per bin to allow spectral
fitting with $\chi^2$-statistic.

The extracted spectra were modelled with XSPEC v12.7.0, using
combinations of an absorbed optically-thin thermal emission (APEC
model, Smith et al. 2001) and power law components.  A single line-of-sight
absorption column $N_H$ was used in all cases, allowed to vary freely.  When
the fit required a very small absorption ($N_H\ll 10^{20}$ cm$^{-2}$),
$N_H$ was frozen at the Galactic column towards NGC 4278 ($1.76\times
10^{20}$ cm$^{-2}$; from CXC tool
COLDEN\footnote{\url{http://cxc.harvard.edu/toolkit/colden.jsp}}).
The spectral modeling was done using the full 0.3--8.0 keV energy
range. In the thermal APEC model, the abundance was fixed to solar,
with the abundance table from Anders \& Grevesse (1989).

\section{Imaging Analysis}

\subsection{The X-ray Morphology}\label{morph}

Figure 1a gives the $4\arcmin \times 4\arcmin$ view of the X-ray
emission of NGC 4278.  The bright nucleus
and the point sources are clearly visible (not removed). There is extended
emission in the inner $30\arcsec$, with a clear elongation along the
northeast (NE; top-left) and the southwest (SW; lower right)
direction.  Figure 1b zooms in to the central $70\arcsec$-across
($\sim$5.5 kpc) region, and Fig.~\ref{f1}c shows the
0.3--2 keV emission in the innermost $20\arcsec$ region.  An
extended feature is present SW of the nucleus, but not seen in other
wavelengths when available images in the literature were examined.

Figure 2a shows the azimuthally averaged radial profile of the
extended X-ray emission in the 0.3--0.9 keV energy band, which
emphasizes the hot ISM presence, since harder emission mostly arises
from unresolved stellar sources as binaries.  We have excluded all
detected sources (as explained above), and centered the concentric
annuli in steps of 5\arcsec\/ on the nucleus, identified as the
optical/IR peak.  The raw profile flattens at a radius of $\sim
70\arcsec$, where the field background starts dominating.  We
therefore used the 100\arcsec--140\arcsec\/ region to measure the
local background and subtract it from the emission, to produce the net
profile.  We further extracted counts from every 10$^{\circ}$-wide
sectors, extending from $2\arcsec$ to $60\arcsec$, to create an
azimuthal distribution for the extended soft X-ray emission; this
procedure identifies sectors that contain brighter emission (Fig. 2b),
and the results were used to guide the choice of regions for the
spectral extraction (see Sect~\ref{sec:spec2}).

In Fig. 2c we show the background subtracted radial profiles of the
extended emission along two directions, the NE--SW sector which
contains the X-ray brighter quadrants towards NE (position angle [P.A.] ranging between
20$^{\rm o}$ and 110$^{\rm o}$, counterclockwise from N) and SW
(P.A. from 180$^{\rm o}$ to 290$^{\rm
  o}$), and the NW-SE sector which contains the NW
(P.A. from 290$^{\rm o}$ to 20$^{\rm o}$) and SE (P.A. from 110$^{\rm
  o}$ to 180$^{\rm o}$) quadrants.  The radial profiles in the optical
I-band from Cappellari et al. (2006) and in the near-IR K band from
the Two Micron All Sky Survey (2MASS) Large Galaxy Atlas (Jarrett et
al. 2003) are also shown for a comparison with the stellar light
distribution. The soft X-ray profile of the NE-SW emission
is clearly higher than that of the NW-SE emission within a radius
$r\sim 20\arcsec$; both emissions follow the optical/IR light
profile at $r>30 \arcsec$, while they are steeper than that, inner of
$r=20\arcsec$. The separate brightness profiles of the brighter NE and
SW quadrants closely follow each other, and are not shown here;
the separate profiles of the fainter NW and SE quadrants are also
consistent with each other, though with lower statistics.

Having established the presence, extent and elongation of the soft
X-ray emission, we next turned to investigate the possible presence of
features in the soft X-ray morphology.
We first modeled the extended soft emission (point source removed and
filled) within a radius of $r=50\arcsec$ with a smooth 1-D
$\beta$-model using CIAO's modeling and fitting package {\tt
  Sherpa}\footnote{\url{http://cxc.harvard.edu/sherpa4.4/}} (Doe et
al. 2007). Clear elongation along the NE-SW direction is seen in the
residual image, as expected given that the profile in this direction
keeps higher than that along the NW-SE (see Fig. 2); thus, a 2-D
(elliptical) $\beta$-model was adopted.  Full description of
the model is available at
\url{http://cxc.harvard.edu/sherpa/ahelp/beta2d.html}. The center of
the model image was fixed at the location of the peak intensity of the
X-ray image ($x_0,y_0$), and the 
modeling included a central point source plus a smooth extended
emission, described by a 2-D Lorentz model with a
varying power law, known as a 2-D beta model:
\begin{equation}
f(x,y) = A*(1+[r(x,y)/r_0]^2)^{-\alpha}
\end{equation}
where $r(x,y) = \sqrt{x_1^2(1-\epsilon)^2 + y_1^2}/(1-\epsilon)$,
$x_1 = (x-x_o)cos{\theta} + (y-y_o)sin{\theta}$ and 
$y_1 = (y-y_o)cos{\theta} - (x-x_o)sin{\theta}$, and $\theta$ is the position angle of the major axis
of the ellipse. 
The model was blurred by a kernel representing the PSF in the image
fitting\footnote{See the CIAO thread 
http://cxc.harvard.edu/sherpa/threads/2dpsf/.}. The best fit
parameter values were $r_0=0.87\arcsec \pm0.05$ for the
core radius, $\theta=261^{o}\pm 5^{o}$, $\epsilon=0.17\pm 0.01$
for the ellipticity, and $\alpha=1.1\pm 0.01$.
Figure 3 shows the ratio image between
the background subtracted soft X-ray emission and the 2D model image
(without including the bright nucleus in the model).
Only features with a minimum significance of $2.5\sigma$ are shown.
The point sources were retained to outline their spatial distribution
with respect to the residual features. Along the NE-SW direction there
are prominent clumps of emission, that are likely responsible for the
excess in the radial profile plots along this direction (Figs. 2b and 2c).
There is also a deficit of emission immediately SW of the nucleus. 
The overall morphology is that of an elongated elliptical shape, with 
an irregular distribution of clumps and deficits within it.

\subsection{Comparison with Multiwavelength Maps}\label{multi}

Figure 4 compares the contours of the extended X-ray emission in the
0.3--0.9 keV band with the HST WFPC2 V band image (Carollo et
al. 1997), $Spitzer$ IRAC 8$\mu$m emission (with stellar continuum
removed; Tang et al. 2011), and the ionized gas maps ([OIII], H$\beta$
emission, [OIII]/H$\beta$, EW$_{[OIII]}$) from the SAURON survey
(Sarzi et al. 2006). In each panel, we also show (1) the directions of
the two-sided pc-scale radio jet, (2) the major axis of the optical
isophote, and the rotation axis of the stars in the inner galactic
region; and (3) the rotation axis for the ionized gas (Sarzi et
al. 2006; see also Morganti et al. 2006). Note how
the isophotes of the stars and of the ionized gas are oriented differently, and 
how stars and ionized gas rotate about two slighly misaligned axes (Sarzi et
al. 2006).  The hot gas is clearly elongated but misaligned with the
optical stellar body of the galaxy, and it follows more closely the
ionized gas distribution and the $Spitzer$ IRAC 8$\mu$m emission.

\section{Spectral Analysis}

\subsection{The X-ray Nucleus}\label{sec:spec1}

The X-ray emission from the nuclear source was studied by Younes et
al. (2010), based on the $Chandra $ pointings prior to March 2010, and
an $XMM-Newton$ pointing of 2004. In their results (see their Table
2), the 0.5--8 keV flux varies from $6\times 10^{-13}$ to
$1.8\times 10^{-12}$ erg cm$^{-2}$ s$^{-1}$.  We have examined these previous
data using a different approach by directly applying the pileup model
of Davis (2001), and we confirm the Younes et al.'s results. We study here the nuclear emission,
focussing on the new data. During the two observations acquired in
2010, the count rate was 0.06 per ACIS frame, and the photon pile-up was
negligible ($<2\%$), in contrast to the high pile-up fraction
($\sim$5\%--20\%) when the nucleus was seen at a higher flux (Younes et
al. 2010).  The nucleus appears to be in its lowest flux state during
the 2010 observations (Tab.~\ref{allflux}). 

The nuclear spectra from the last two observations were extracted from
a $2\arcsec$ radius circular aperture, and were fitted jointly
(Fig. 5).  An absorbed power-law model gave no acceptable fit (reduced
$\chi^2=2.8$), in contrast with the highest flux state case (ObsID
4741), where a simple power law with photon index $\Gamma=2$ could fit
the nuclear spectrum.  The 0.3--2 keV band shows residuals around
$\sim $1 keV, strongly suggesting the presence of emission from hot
gas.  The addition of a thermal component ($APEC$ model)
takes into account these residuals in the soft band, and
significantly improves the fit quality (at $>99.9\%$ confidence level,
as established by the F-test). The best fit temperature is $kT=0.75\pm
0.05$ keV and the power law $\Gamma=2.31\pm 0.20$.  The quoted errors
are 90\% confidence intervals for one interesting parameter.  The
observed total nuclear flux is $F_{0.5-8
  keV}=1.02_{-0.17}^{+0.11}\times 10^{-13}$ erg cm$^{-2}$ s$^{-1}$,
and $F_{0.5-8 keV}=1.1\pm 0.2\times 10^{-13}$ erg cm$^{-2}$ s$^{-1}$
after correction for absorption ($N_H=4.18\pm 3.13\times 10^{20}$). We
also attempted to fit the two observations individually, and found
$F_{0.5-8 keV}=1.17\pm 0.05\times 10^{-13}$ erg cm$^{-2}$ s$^{-1}$ for
ObsId 11269 and $F_{0.5-8 keV}=0.95\pm 0.08\times 10^{-13}$ erg
cm$^{-2}$ s$^{-1}$ for ObsId 12124 (both absorption corrected); the
nuclear flux is slightly lower in the latter observation.  In any
case, the flux was a factor of 18 lower than the highest flux reported in
Younes et al. (2010).  The comparison with the Younes et al. (2010)
results is summarized in Table~\ref{allflux}.  The nuclear luminosity
due to the power law component is $L_{0.5-8\,{\rm keV}}=2.5\times
10^{39}$ erg s$^{-1}$; Tab.~\ref{nuc_spec} gives the luminosities of
the thermal and power law  components in various bands.

The thermal component alone has $L_{0.5-2keV}=9.9\times 10^{38}$ erg
s$^{-1}$ (absorption corrected), and accounts for 40\% of the total
0.5--2 keV emission within the $R=2\arcsec$ region
($L_{0.5-2keV}=2.5\times 10^{39}$ erg s$^{-1}$, absorption corrected;
Tab.~\ref{nuc_spec}). We can rule out the possibility that a
significant fraction of the thermal emission at the nucleus position
is originated from the combined contribution from the AB+CV stellar
population (e.g., Boroson et al. 2011; see also Sect.~\ref{sec:spec2}
below), since the latter is estimated to account for only $\sim$0.5\%
of the 0.5--2 keV hot gas emission in the extraction region. This
estimate is based on the 2MASS K-band image, which gives a K-band
luminosity of $1.0\times 10^9 L_{K,\odot}$ for the region used for the
extraction of the nuclear spectrum (for $M_{K,\odot}=3.3$ mag).
Adopting the $L_{0.5-2{\rm keV}}/L_K=4.4\times 10^{27}$ erg
s$^{-1}/L_{K,\odot}$ conversion relation for the AB+CV emission
(eq. A1 in Boroson et al. 2011), this gives a stellar AB+CV luminosity
of $L_{0.5-2{\rm keV}}=4\times 10^{36}$ erg s$^{-1}$.

\subsection{The extended emission}\label{sec:spec2}

We derived the temperature and density profiles of the hot gas using
all available $Chandra$ ObsIDs.  Due to statistics, to obtain enough
counts, we analyzed the combined spectra extracted from radial bins
along two directions, the NE--SW and the NW--SE sectors defined in
Sect. 3.1.  The sectors were divided in consecutive annular regions,
centered at the nucleus, and selected to have $\sim$1000 (background
subtracted) counts in the spectra (inner-to-outer radii given in
Tab.~\ref{depro}).  The temperature and density of the innermost
2\arcsec\/ radius region were derived using the last two ObsIDs
(11269,12124) together, to avoid as much as possible contamination
from the nucleus.

In addition to the sky background, it is necessary to account for the
stellar contribution to the X-ray emission. We adopt the spectral
model derived for M32 to account for the unresolved AB+CV emission
(Boroson et al. 2011; Li, Z. et al. 2011), normalized by scaling it to
the K-band light enclosed within the spectral extraction regions. The
AB+CV model is characterized by a combination of two thermal plasma
emission components, with solar abundance and fixed temperatures of
0.4 keV and 4.6 keV, which are dominated by the emission of ABs and
CVs, respectively (see Li, Z. et al. 2011 for details). An additional
power law component, with a photon index of $\Gamma =1.8$ and a normalization
free to vary, is adopted to account for the residual contribution from
low mass X-ray binaries (LMXBs; e.g., Kim et al. 2009).  To account
for PSF scattering of the nuclear emission, the best fit nuclear
spectrum was included as a model component, with a scaling factor
fixed to the PSF wing fraction (as determined from ChaRT and MARX
simulations) in the spectral extraction region; this component is
negligible ($\ll 1\%$) in all regions except the first annulus.  We
then characterize the diffuse gas by an APEC model component (with
solar metal abundance).

Assuming a spherical geometry, we used the $projct$ model in XSPEC to
perform a spectral deprojection.  Before fitting, to account for the
fact that only a fraction of the area is covered due to source
removal, the ARFs generated for the annular regions were rescaled
according to the missing area from source masking.
The resulting best-fit spectral parameters are summarized in
Tab.~\ref{depro}. The temperature is consistent with the $0.32\pm
0.02$ keV value reported in Boroson et al. (2011) at the outer radii, but
considerably hotter in the nuclear region and in the first annulus
(2\arcsec--4\arcsec\/), in all directions (Fig. 6). Without deprojection, this
inner hot component with a similar temperature is found again at a
significant level, thus it is not due to the $projct$ model.

Given the peculiar and unusual result of a significantly hotter
thermal emission at the nucleus and in the surrounding $2\arcsec -
4\arcsec$ annulus, we next investigated whether the high temperature
is a real feature of the hot gas, or could be due to an underestimate
of the nuclear
contribution.  We focus on ObsID 11269 because of its depth and well
characterized nuclear emission (pile-up free) in a single observation.
We performed ChaRT+MARX simulations and found that the PSF wing may
scatter 2\% of the nuclear APEC component to the 2\arcsec--4\arcsec\/
radial bin (or a 0.5--2 keV flux of $6.0\pm 1.2 \times 10^{-16}$ erg
cm$^{-2}$ s$^{-1}$).  The flux of the thermal component is $9\times
10^{-15}$ erg cm$^{-2}$ s$^{-1}$, which requires the PSF scattering
fraction of the nuclear emission to be unrealistically underestimated
($30\%$ instead of $2\%$).  Moreover, the contribution from unresolved
stellar sources is an order of magnitude lower than the $APEC$
component.  Thus we conclude that the hot thermal component is
significant, and arises from gas at a higher temperature in the inner
$\sim $few hundreds pc.

The deprojected density in each spherical shell is shown in
Fig. 7. Densities, along with other physical properties of interest
for the hot gas, are presented in Tab.~\ref{phys}.  We accounted for
the fact that only a fraction of the area within a given annulus was
covered, due to source removal.  The electron number density $n_e$ of
the hot gas was derived from the emission measure ($\int n_e^2 dV$) of
the APEC component given by the spectral fit of each extraction
region.  A spherical geometry was assumed, and the volume covered by
the quadrants was properly calculated using the apex angle of the two
quadrants.

\section{Summary of the observational results and discussion}

A deep (579 ks) $Chandra$ observation of the elliptical galaxy NGC4278,
including two pointings during year 2010 and six pointings prior to
2010, has been analyzed with the following main results:

$\bullet $ The nuclear emission dominates the X-ray image, but its
0.5--8 keV flux in 2010 was the lowest among all pointings,
and $\sim 18$ times lower than at its brightest state seen with $Chandra$
(in 2005). At the nucleus, within $r=2^{\prime\prime}=156$ pc, a power law
spectral component (with $\Gamma=2.31\pm 0.20$, consistent with AGN
emission) and a thermal component (with $kT=0.75\pm 0.05$ keV)
coexist, respectively with an average $L_{0.5-8\,{\rm keV}}=2.5\times
10^{39}$ erg s$^{-1}$ during year 2010, and with $L_{0.5-8\,{\rm keV}}=1.0\times 10^{39}$
erg s$^{-1}$.

$\bullet $ After merging data from all pointings, hot gas is detected
out to a radius of $\sim 5$ kpc, with a total $L_{0.5-8\,{\rm keV}}=2.4\times
10^{39}$ erg s$^{-1}$. On the galactic scale, the hot gas shows an
elliptical shape, elongated in the NE-SW direction; its distribution
is clearly different from that of the stellar component of the galaxy,
having a flatter shape and a different orientation. The
hot gas seems to follow the distribution of the ionized gas, that
resides in a rotating, inner disk, and of the warm dust emission
detected with $Spitzer$.

$\bullet $ The image ratio with the best-fit 2-D modeling of the
surface brightness (i.e., the
unsharp masking, Fig. 3) shows that the gas distribution
surrounding the bright nucleus is not smooth, but includes
regions where the brightness is lower than the average, 
and various clumps. In particular, two larger clumps are present NE
and SW of the nucleus, at a distance of $\sim 10\arcsec$. These
regions of enhancement of the brightness are not likely to be 
concentrations of stellar sources, being clearly extended; the
spectral analysis of the diffuse emission also indicates thermal
emission from hot gas, and not from fainter LMXBs.

$\bullet $ The 3D temperature and density profiles derived from
deprojection along the NE-SW and NW-SE directions are
consistent with each other at all radii, except for the temperature in
the annulus surrounding the nucleus, that is lower in the direction
of the elongation (NE-SW).  The hot gas temperature stays at 0.6--0.8
keV out to a radius of $\sim 300$ pc, and drops sharply to $\sim 0.3$
keV outside, keeping a constant (or slowly decreasing) value out
to $\sim 5$ kpc.  The density profile shows a smooth decrease
throughout the region examined.

The above results prompt the following questions: what is the origin
of the nuclear emission, and of the hotter gas at the galactic center?
what produces the hot gas elongation? is there a relationship between
the hot and the warm gas that seem to follow the same projected
distribution ?  Taking also advantage of the radio information, we now
investigate the hot gas origin and evolution, and the activity cycle,
in this medium-mass elliptical galaxy; systems of this mass have not
been investigated in detail previously. We examine in turn the origin
of the nuclear emission in the next Sect.~\ref{nuc}, that of the
central hot gas in Sect.~\ref{spike}, and in Sect.~\ref{cold} the
relationship between the hot, warm, and cold gas phases.

\subsection{The nuclear emission}\label{nuc}

The X-ray spectral shape of the low luminosity AGN during the most
recent March 2010 pointings is consistent with that determined from
pointings made between 2005 and 2007 by $Chandra$, and by $XMM$-Newton
in 2004 (Tab. 2): during these previous brighter states, the average
photon index was $\Gamma=2.2^{+0.1}_{-0.2}$, affected by small
intrinsic absorption ($N_H<6.7\times 10^{20}$ cm$^{-2}$; Younes et
al. 2010).  The source was brightest in 2004, and decreased by a
factor of $5.7$ by 2007; this trend of decrease in flux continued
through 2010 (Sect.~\ref{sec:spec1}).  Based on the spectral
energy distribution from radio to X-rays, it was suggested that at low
X-ray flux the nuclear emission is more typical of
LINERs (and possibly originates in a RIAF and/or a jet), whereas at
high X-ray flux it resembles more that of a Seyfert (Younes et
al. 2010). The nucleus in 2010 had $L_{0.5-8\,{\rm
    keV}}/L_{Edd}=5.9\times 10^{-8}$, where the Eddington luminosity
$L_{Edd}=4.25\times 10^{46}$ erg s$^{-1}$ for the MBH mass in
Tab.~\ref{tab1}. The bolometric correction ($L_{bol}/L_X$) for low
luminosity AGNs is believed to be lower than the value of standard
AGNs ($\sim 30$, Elvis et al. 1994), where the accretion disk
dominates the emission; for example, Ho (2009) suggested that
$L_{bol}/L_{2-10\, {\rm keV}}\approx 8$ for low luminosity AGNs, and,
for a large sample including nuclei with $L_{bol}/L_{Edd}\lsim 0.1$, a
median $L_{bol}/L_{2-10\, {\rm keV}}\approx 15.8$.  Therefore the AGN in NGC4278
is a very sub-Eddington radiator  (Tab.~\ref{nuc_spec}
gives the 2--10 keV nuclear luminosity), and its low luminosity could be the
result of a low radiative efficiency, provided that the mass accretion
rate is low.  Thanks to the large dimming of the nucleus in 2010, we
could derive the gas properties close to the MBH, a result used in
what follows to derive an estimate of the mass accretion rate, and
then to discuss the accretion modalities.

For material accreting on the MBH at rate $\dot M$, such that $\dot
m=\dot M/\dot M_{\rm Edd}<<0.01$ [where ${\dot M_{\rm Edd}}= 22
M_{BH}(10^9 M_{\odot})$ $M_{\odot}$ yr$^{-1}$], the accretion flow can
become a RIAF, with an efficiency for producing radiation of $\epsilon
\sim 10\dot m $ (Narayan \& Yi 1995); the expected $L_{bol}$ is then
$\sim 10 \dot m \dot M c^2$.  We can adopt for $\dot M$ the mass
accretion rate given by the steady and spherically symmetrical Bondi
(1952) solution, $\dot M_B$ (e.g., Loewenstein et al. 2001, Di Matteo
et al. 2003, Pellegrini 2005). An accretion rate $\lsim \dot M_B$
enters also in the viscous rotating analog of the Bondi treatment
represented by the RIAF models (Quataert 2003, Narayan \& Fabian
2011).  $\dot M_B$ is given by $\dot M_B=\pi G^2 M_{BH}^2
{\rho_{\infty}\over c_{s,\infty}^3}\left[ {2\over
    5-3\gamma}\right]^{(5-3\gamma)/2(\gamma-1)} $, where $\gamma$ is
the polytropic index [$\gamma =1$ (isothermal) to 5/3 (adiabatic)],
$c_s$ is the sound speed, and ``$\infty $'' refers to the ambient
conditions (e.g., Frank et al. 2002).  Ideally, one should insert in
thsi formula the gas density and temperature at the accretion radius
$r_{acc} =2GM_{BH}/c_{s,\infty}^2$, where the dynamics of the gas
start to be dominated by the potential of the MBH. In practice, one
uses fiducial temperature and density for the circumnuclear region,
determined as close as possible to the MBH.
For the values for the
nuclear region in Tabs.~\ref{depro} and~\ref{phys}, one obtains
$r_{acc}=15-25$ pc, and $\dot M_{B}=(0.5-5)\times 10^{-3}\, M_{\odot}$
yr$^{-1}$ (for $\gamma=5/3-1$); the corresponding $\dot
m_{B}=(0.7-7)\times 10^{-4}$ is very low, within the RIAF regime. This
$\dot M_{B}$ is based on gas properties that are an average for a
sphere of 156 pc radius ($\sim 6r_{acc}$), thus its value should be
taken with some caution; for example, this $\dot M_B$ may
underestimate the true mass accretion rate at $r_{acc}$ if the density
rises steeply towards $r_{acc}$ (but this seems not to be the case; see
the modeling of Sect.~\ref{nofeed}).

Using an average $\gamma=4/3$, then $\dot M_{B}=2.1\times 10^{-3} \,
M_{\odot}$ yr$^{-1}$, and $L_{bol}= 10 \dot m_B \dot M_{B} c^2\sim
3.3\times 10^{41}$ erg s$^{-1}$ (while a standard accretion disc would
have a radiative output of $L_{acc}=0.1\dot M_{B} c^2\sim 1.2\times
10^{43}$ erg s$^{-1}$); for the whole range of $\dot M_B$,
$L_{bol}\sim (0.02-2)\times 10^{42}$ erg s$^{-1}$.  Adopting a
correction factor appropriate for the spectral energy distribution of
a RIAF, i.e., $L_{0.5-8 {\rm keV}}\lsim 0.15 L_{bol}$ (Mahadevan
1997), for the average $\dot M_{B}=2.1\times 10^{-3} \, M_{\odot}$
yr$^{-1}$ one expects $L_{0.5-8 keV}\lsim 5\times 10^{40}$ erg
s$^{-1}$, that is $\lsim 20$ times larger than the observed $L_{0.5-8
  keV}=2.5\times 10^{39}$ erg s$^{-1}$ (the expected 
$L_{0.5-8  keV}\lsim 2.8\times 10^{39}$ erg s$^{-1}$, for the lowest estimate
of $\dot M_B$, and $\lsim 3\times 10^{41}$ erg s$^{-1}$ for the
largest). Reductions of the mass accretion rate on the way to the MBH
have often been claimed for RIAFs, since they include solutions where
little of the mass available at large radii is accreted on the MBH due
to outflows or convective motions (Blandford \& Begelman 1999; Stone,
Pringle, \& Begelman 1999; Igumenshchev, Narayan, \& Abramowicz 2003).
Another source of reduction is given by the possibility that the gas
has non-negligible angular momentum beyond the Bondi radius (Proga \&
Begelman 2003; Narayan \& Fabian 2011). The latter authors calculated
the rate at which mass accretes on to a MBH from rotating gas, for
RIAFs in galactic nuclei, and found that $\dot M\sim (0.3-1)\dot
M_{B}$, for a plausible viscosity parameter value ($\alpha
=0.1$). Large reductions of $\dot M$ with respect to $\dot M_B$ are
not required here; for example, if $\dot M\sim 0.3\dot M_{B}$,
accounting for only the effect of rotation, then $L_{bol}$ would
decrease by a factor of $\sim 10$, and the predicted $L_{0.5-8 {\rm
    keV}}$ for the average $\dot M_{B}$ would be $\lsim 5\times
10^{39}$ erg s$^{-1}$, close to that observed. The largest possible
reduction in $\dot M_B$ still reproducing the observed luminosity is
of a factor of $\sim 10$ (calculated for the largest estimate of $\dot
M_B$).

The accretion process at the nucleus of NGC4278 has also been studied
recently by modelling the radio-to-X-ray emission with an inner RIAF, an
outer truncated thin accretion disk, and a jet, with the possibility
of a radially varying mass accretion rate, within the RIAF region, to
account for outflows (Nemmen et al. 2011).  The radio was explained
with a jet origin, the 100$\mu m$ to 1$\mu m$ emission
required an outer accretion disk truncated at 
30--100 Schwarzschild radii, and the X-rays could be equally
well reproduced mostly by the RIAF (from inverse Compton scattering of
synchrotron photons in the flow) or mostly by the jet (from
synchrotron photons from the jet; see also Yuan \& Cui 2005).  The best fit models, respectively
for the RIAF-dominated and the jet-dominated cases, have
an accretion rate at the outer disk of\footnote{$\dot m$ is the Eddington-scaled mass accretion rate.}  $\dot
m_{out}=7$ or $4\times 10^{-4}$, and an accretion rate on the MBH of
$\dot m_{in}=4.4$ or $0.9\times 10^{-4}$.  These $\dot m_{out}$ values
are within the $\dot m_{B}$ range estimated here, and $\dot m_{in}$ is
not too different from $\dot m_{out}$; for the case of the X-rays
coming mostly from the RIAF, $\dot m_{out}=1.6 \dot m_{in}$, a result
similar to what found above for the small reduction (if any) of 
$\dot m_B$ to reproduce the observed $L_{0.5-8{\rm keV}}$. In fact, 
the $\dot m_B$ from our analysis, for $\gamma=1.5$ as adopted by Nemmen et
al., produces $L_{0.5-8{\rm keV}}\lsim 1.4\times 10^{40}$ erg s$^{-1}$
if ending in a RIAF, and it should then be reduced by  $\lsim 2.3$
times to reproduce the observed luminosity.

In conclusion, the hot gas observed with $Chandra$ near the MBH has
density and temperature consistent with the idea that the nuclear
luminosity comes from a RIAF, with a $\dot M$ close to that estimated
with the Bondi formula. If all the nuclear X-rays are to come from the
accretion flow, $\dot M_B$ can be reduced by a factor expected in case
of rotation, still reproducing the observed luminosity; the latter
constrains the reduction to be not larger than a factor of $\sim 10$.  If
the X-rays come mostly from a nuclear jet, the accretion flow
luminosity must be lower than that observed, and then a more important
reduction for $\dot M_B$ is allowed for, as can be achieved when
outflows or convection are added. A further important constraint on
how much mass must be accreting is given by the total energy output
from accretion, as will be discusssed in Sect. 5.2.3.

\subsection{The origin of the central hotter gas}\label{spike}

In a galaxy of the size of NGC4278, given standard assumptions
concerning the type Ia supernova (SNIa) heating, the mass losses from
evolving stars are expected to originate an outflow on the galactic
scale (see also Sect.~\ref{nofeed} below); this agrees with the low
observed hot gas content, of a few$\times 10^7$ M$_{\odot}$
(Tab.~\ref{phys}). The intriguing central peak in temperature,
instead, is a new finding, that could be produced by the following
causes: 1) the MBH gravity field, 2) a nuclear outburst triggered by a
past high accretion rate phase (i.e., in the AGN-mode), 3) hot
accretion during a quiescent, low accretion rate state, 4) kinetic
heating from the jet.

In the following, these hypotheses are examined in turn.  The first
one is studied with hydrodynamical simulations
(Sect.~\ref{nofeed}); the next two are examined in light of the
predictions of previous results from numerical simulations
(Sect.~\ref{feed}); the last one is investigated through energetic
calculations also exploiting the radio information (Sect.~\ref{jet}).

\subsubsection{Hot gas evolution without AGN feedback}\label{nofeed}

We performed hydrodynamical simulations for the evolution of the stellar
mass losses, for a detailed galaxy model built for NGC4278, without
feedback from the MBH; the aim is to establish what can
and what cannot be explained by this basic model.  The
underlying 3-component galaxy mass model is made by the superposition
of a stellar and a dark distributions, {\it plus a central MBH}, of
mass fixed at that in Tab.~\ref{tab1}.  All other details regarding
the galaxy model and the simulations are given in the Appendix.

The resulting gas flow on the galactic scale is an outflow driven by
SNIa's heating during the entire evolution; a small accretion region
at the center is always present, due to the cuspy mass profile,
typical of elliptical galaxies (e.g., Pellegrini 2012).  With time
increasing, the SNIa's specific heating ($L_{SN}/\dot M_*\propto
t^{0.2}$, where $L_{SN}$ is the SNIa's heating rate and $\dot M_*$ is
the stellar mass loss rate) slowly increases; this produces a decrease
of the inflow velocity and of the radius of the inflowing region, that
goes down to $\sim 30$ pc at the present epoch.  At the center, the
gas density keeps decreasing and the temperature increasing with time
(Fig.~\ref{hyd1} in the Appendix); mass flows through the innermost
gridpoint (5 pc) at a decreasing rate, reaching $\dot M\sim 0.003
M_{\odot}$ yr$^{-1}$ at the present epoch.  Even though the density
keeps rising towards the center in the model, this value of $\dot M$
is within the range derived for $\dot M_B$ in
Sect.~\ref{nuc}  (at its upper end).

The temperature and density profiles at an age consistent with that of
the stellar population of NGC4278 ($10.7\pm 2.14$, Terlevich \& Forbes
2002) are shown in Fig.~\ref{hyd}.  Initially, the temperature drops
towards the center, with a central value of $\sim 0.55$ keV at an age
of $\sim 10$ Gyr; this value is an average calculated for a central
sphere of radius of $2\arcsec$, as in the observations, and includes the
contribution of a very inner region where the central MBH creates a
spike in the gas injection temperature (see Fig.~\ref{hyd}). Later,
the central temperature becomes larger ($\sim 0.85$ keV), and the
temperature profile decreases smoothly out of $2\arcsec$
(Fig.~\ref{hyd}, dotted line), without a sharp drop as instead
observed.  The model temperature is then larger than observed at the
center, and agrees with that observed in the surrounding annulus of
radii $2^{\prime\prime}-4^{\prime\prime}$ only for the NW-SE quadrant;
therefore, the observed temperature cannot be reproduced by the models
as a long-lasting feature\footnote{A temperature profile within $\sim
  400$ pc similar to the observed one is produced between the two
  epochs shown in Fig.~\ref{hyd}, but for a very brief time, lasting
  $\lsim 10^7$ yr.}.  The model density profile is in reasonable
agreement with the $Chandra$ one at 10 Gyr, but it is lower than that, inwards
of $\sim 400$ pc, for later epochs (Fig.~\ref{hyd}).  The MBH clearly
heats the gas in its surroundings, through the combined effects of its
gravitational field (causing gas compression) and of the increase in
the stellar heating, due to the stellar velocity dispersion
enhancement produced by the MBH within its sphere of
influence\footnote{The sphere of influence of the MBH has a radius of
  $r_{BH}=GM_{BH}/\sigma_0^2= 23$ pc, for the $M_{BH}$ and central $\sigma$
  values in Tab.~\ref{tab1}.}  (see the Appendix and its
Fig.~\ref{mass}).  This heating, though, keeps the gas temperature
very high on a much smaller scale than observed 
(Fig.~\ref{hyd}).  Evidently, the gas properties close to the center
are established by different (additional) phenomena, with respect to
those simply connected with the MBH gravity, the SNIa's heating, and
mass input from stellar mass losses. This analysis also shows
that accretion is possible even for gas with a central temperature as large as observed.

Another result of this investigation is that the observed temperature is
lower than in the model, outwards of a $\sim 0.5$ kpc radius, and
consistent just with the temperature expected from the thermalization
of the stellar motions ($T_{\sigma}$; Fig.~\ref{hyd}).  Note that the
stellar heating could be lowered by at most $\sim 20$\%, for different
mass models still consistent with the observational constraints. When
coupled to the fact that also the model density tends to be lower than
observed outside $\sim 1$ kpc, possibly indicating too much
degassing, all this could point to a lower efficiency of the SNIa's
energy mixing process than standardly adopted.  In fact the fraction
of the SNIa's kinetic energy that is turned into heat is uncertain,
and may depend on the environment surrounding the expanding supernova
remnants (e.g., Tang et al. 2009, Pellegrini 2011).  In massive, hot
gas-rich early type galaxies, SNIa's bubbles should disrupt and share
their energy with the local gas within $\sim 3\times 10^6$ yr (Mathews
1990); in less massive spheroids in a global wind, instead, 3D
hydrodynamical simulations of discrete SNIa's heating suggest a
non-uniform thermalization of the SNIa's energy, with overheated gas
advected outwards, carrying a large fraction of the SNIa energy with
it (Tang et al. 2009).  Another possibility is that the gas is cooled
by the presence of other colder phases in the ISM (see
Sect.~\ref{cold}).

\subsubsection{Hot gas evolution with AGN-like outbursts}\label{feed}

In the recent past NGC4278 could have experienced a nuclear outburst
with a high $\dot m$, and the high central value of the gas temperature
could be a remnant of that episode.  Indeed, the inner $\sim 4\arcsec$
seem to contain a hot bubble, a region that appears to be
overpressured with respect to its surroundings. Also, the absence of
HI absorption against the nucleus (van Gorkom et al. 1989) suggests that
the region around the center has been cleared of neutral gas; the
$HST$ WFPC2 images show that in the inner $\sim 2\arcsec$ diameter 
the dust lines are absent, and a pointlike bright nucleus is
seen (Lauer et al. 2005). All this would be consistent with an inner
region where the cooler gas has been removed, thus there is no dust
and no HI absorption around the nucleus (consistent with the spectral
results, Sect.~\ref{sec:spec1}).  The effects of AGN feedback
typical of accretion at high $\dot m$ have been investigated in detail
previously with spherically symmetric hydrodynamical simulations
(Ciotti et al. 2010).  During outbursts, very hot central gas is
created (within $\sim 100$ pc), at a temperature that can reach up to
a few keV; this central bubble is surrounded by gas at much lower
temperature ($kT$ drops down to 0.3--0.4 keV at $\sim 1$ kpc;
Pellegrini et al. 2012), which resembles the observed trend in
NGC4278. However, there are two caveats: the first is that the hot
bubble lasts for a short time ($< 10^8$ yr), since it cools quickly
and, soon after it is created, the accretion rate drops; furthermore,
the outburst produces a low density region surrounding the nucleus,
with an almost flat 0.3--2 keV brightness profile, extending to $\sim
$ a few kpc and lasting for $\sim 10^8$ yr (Pellegrini et al. 2012),
and this large plateau in the brightness profile is not observed here
(Fig. 2c). The second caveat is that high $\dot m$ outbursts are
coupled with starformation in the galactic central region, but NGC4278
shows a uniformely old stellar population, without signs of current or
recent starformation (Shapiro et al. 2010, Kuntschner et
al. 2010). All this supports the idea that nuclear outbursts at high
$\dot m$ are well confined in the past.  In fact, an isolated galaxy
of the mass of NGC4278 is expected to have a very small duty-cycle for
the phases of accretion at high $\dot m$, over the past few Gyr, with
the last bursting episode confined to a few Gyr ago (Ciotti \&
Ostriker 2012).

As these simulations with accretion at high $\dot m$ show, after an
outburst is terminated, a new sub-Eddington accretion phase
establishes, where the nucleus is radiatively quiescent ($\dot m$ is
in the range of the RIAF regime). The galactic center, almost empty of
gas, is replenished with the newly injected stellar mass losses, that
feed the accreting flow and are kept hot by the MBH via compression,
stellar heating, and the energy output from accretion.  The gas
behavior during this phase is similar to what is described in
Sect.~\ref{nofeed}, and the temperature profile resembles that
obtained without feedback at an epoch of 10.9 Gyr in Fig.~\ref{hyd}
(Pellegrini et al. 2012); there is though less hot gas (at the same
age).  Therefore, even a post-outburst radiatively quiescent phase
is not likely to explain the observed gas properties.  The models considered
in this Section include feedback expected from high $\dot m$
accretion, that is in radiative and mechanical form, with the latter
from AGN winds only. At low $\dot m<<0.01$ the bulk of the accretion
output is thought to be mostly mechanical, emerging in the form of
jet/outflows rather than radiation (e.g., Merloni \& Heinz
2008). Given the presence of a pc-scale jet in NGC4278, we examine in
the next Sect.~\ref{jet} whether a central hot region
can be produced by mechanical heating.
Note that the different temperature but the very similar density observed in the
$2\arcsec-4\arcsec$ annulus, in the two NW-SE and NE-SW directions
(Figs. 6 and 7), implies a variation in the thermal gas pressure from
one quadrant to the next, at this fixed distance from the center: the
pressure is larger perpendicular to the direction of the gas
elongation (where the temperature is larger). The heating may have
taken place along this preferential direction (the NW-SE one), that is
also close to the jet directions (Fig. 4).

\subsubsection{Jet heating (and confinement)}\label{jet}

VLBA observations of the NGC4278 nucleus with a resolution of a few
mas ($\sim 0.1$ pc) at 5 and 8.4 GHz revealed synchrotron radio emission
from a two-sided structure, made of two symmetric S-shaped 
jets emerging from a flat-spectrum core (Giroletti et al. 2005; see
also Falcke et al. 2000). In total, the two-sided jet extends over
$\sim 45$ mas, that is 3.5 pc. The distance reached by the
currently observed jets is then smaller than the accretion radius
($r_{acc}\sim 15-25$ pc, Sect.~\ref{nuc}), and much smaller 
than the size of the observed hot region (that extends
out to $4^{\prime\prime}$, in the NW-SE direction).

The mildly relativistic velocity for the jet ($v_j/c\approx
0.76$) implies an epoch of jet ejection that is 10--100 yrs prior to
the observation (Giroletti et al. 2005). These authors suggest that
this source does not evolve into a kiloparsec scale radio galaxy, but
rather ejects components that soon disrupt, without being able to
travel long distances and form radio lobes.  The lack of hot spots,
that are instead prevalent in higher power compact symmetric objects
(CSOs), indicates that the relatively low velocity jets may not be
able to bore through the ISM and escape.  The small size of the radio
source was ascribed to a low-power central engine, which cannot create
highly relativistic jets, possibly combined with the interaction with an interstellar
medium.  We address then the following questions: what is the
energy flux for the jets? Is it likely that they are frustrated by the
hot ISM in the nuclear region? Is it possible that (a fraction of) the
jet energy has been deposited in this region, creating the hot central
gas?

The average thermal pressure of the X-ray gas for the central
$2^{\prime\prime}$ radius sphere is $5.8\times 10^{-10}$ dyn cm$^{-2}$
(Tab.~\ref{phys}), a value comparable to the pressure in the
atmospheres of hot gas rich giant ellipticals, where the hot ISM often
confines radio sources (e.g., Allen et al. 2006, Cavagnolo et
al. 2010).  We can estimate the minimum pressure for the jets of
NGC4278 from synchrotron theory, assuming energy equipartition between
particles and magnetic field (e.g., O'Dea \& Owen 1987). The spectral
index of $a=-0.54$ ($S\propto \nu^{a}$) gives a good description of
the emission over the whole radio band (VLA measurements are available
from 74 MHz to 22 GHz), with a possible flattening below 408 MHz
(Giroletti et al. 2005).  Calculating a 74MHz--22GHz radio luminosity
for the jets of $6.2\times 10^{38}$ erg s$^{-1}$, and assuming a
volume emitting region for the jet that is a cylinder of height 45 mas
and radius $r_j=4$ mas, with filling factor $\phi =1$, gives
$p_{min}\sim 2\times 10^{-6}$ dyn cm$^{-2}$.  Then the jets can be
thermally confined only if within few parsecs the thermal pressure
becomes much larger than the average for the central
$2^{\prime\prime}$ (=156 pc). However, this $p_{min}$ value is likely
uncertain (e.g., due to the unknown source volume).  An alternative
way to estimate the jet internal pressure $p_{j}$ is via the energy
flux carried by a jet, that is via the jet power $P_{jet}\approx 4\pi
r_j^2 v_j \gamma_j^2 p_j$, with $\gamma_j$ the jet Lorentz factor
(e.g., Owen, Eilek, \& Kassim 2000).  $P_{jet}$ can be estimated from
its relationship with the 5GHz radio core luminosity (Merloni \& Heinz
2007); for an average $P_{5GHz}=4.5\times 10^{37}$ erg s$^{-1}$ for the core
component resolved by the VLBA (Giroletti et al. 2005), the relation
gives $P_{jet}=2.5\times 10^{42}$ erg s$^{-1}$. Cavagnolo et al. 2010
also presented relationships between the total synchrotron power, at
200--400 MHz, and the mechanical power of jets, measured
from the cavity power. For a total 
$P_{200-400\,{\rm MHz}}=4.7\times 10^{37}$ erg s$^{-1}$, then 
$P_{jet}=1.1\times 10^{42}$ erg s$^{-1}$.
The jet power, again inferred from the cavities observed in the hot
ISM of nearby radio ellipticals, correlates also with the accretion
power $L_{acc}=0.1\dot M_{B} c^2$ (Allen et al. 2006). Inserting in
this correlation the average $\dot M_{B}\sim 2.1\times 10^{-3}$
M$_{\odot}$ yr$^{-1}$ of Sect.~\ref{nuc}, one finds $P_{jet}=1.8\times
10^{42}$ erg s$^{-1}$, in good agreement with the values given by the
scalings with the radio luminosity. Note that
these $P_{jet}$ estimates imply a jet production efficiency
$\epsilon_{jet}=P_{jet}/\dot M_{B} c^2=0.01-0.03$, that is $P_{jet}$
is a significant fraction of $L_{acc}$. Unless $P_{jet}$ is largely
overestimated, mass needs then to be accreted at approximately the
Bondi rate, derived for the gas properties close to the accretion
radius; large reductions in the rate along the way to the MBH are not
allowed (see also Allen et al. 2006). This completes the findings of
Sect.~\ref{nuc} abouth the origin of the nuclear $L_X$, excluding the
possibility that $\dot M_{B}$ is reduced by orders of magnitude.

Adopting $P_{jet}\sim (1-2)\times 10^{42}$ erg s$^{-1}$,
$\gamma_j=1.5$ and $v_j=0.76c$ (Giroletti et al. 2005), then $p_j\sim
(1.6-3.3)\times 10^{-6}$ dyn cm$^{-2}$.  To overcome this pressure,
the hot ISM should become much hotter and denser within few pc from
the MBH (the product $nT$ should increase by a factor of $\gsim 10^3$
above the average within $2^{\prime\prime}$). Whether this is indeed
verified remains beyond testability with $Chandra$, but could be
feasible, within a factor of few; for example, in a Bondi flow with
$\gamma=4/3$, the thermal pressure is $\propto r^{-2}$, and then can
increase by a factor of $\sim 10^3$ moving inward by a factor of $\sim
30$ in radius; in a RIAF, the behavior of the pressure is similar
(Narayan \& Fabian 2011). Other cooler gas phases may contribute to
the confinement of the jets, as cold gas (e.g., Emonts et al. 2010); however,
there is no evidence for them towards the nucleus, as discussed in
Sect.~\ref{feed}.  Alternatively, $P_{jet}$ derived from the
correlations above may be overestimated. If confinement is not
feasible, the currently observed jets are really very young and just
making their way through the hot ISM.  Note, though, that extended
radio emission out of the nucleus has not been detected, and that going
down to the lowest frequency observed ($\sim 0.1$ GHz), where ``old''
radio lobes should contribute if present producing an excess of
emission, the radio spectrum does not steepen, but rather, if any, it
flattens (e.g., Giroletti et al. 2005).  From the relationship between
a critical frequency above which the spectrum steepens due to ageing,
that in this case is $< 1$ GHz, and the electron age, one derives that
old, extended emission must date back to at least a few$\times 10^8$
yr (for a magnetic field of a few $\mu$G, as reasonable for an old
radio lobe; e.g., Feretti \& Giovannini 2008).  Past radio activity
seems then to have always been confined within the nuclear scale, or
it took place more than a few$\times 10^{8}$ yr ago.

A jet could have heated the central gas via dissipation of the kinetic
energy of shocks produced in the circumnuclear region.  The heating by
the jet of a region much larger than its currently observed extension
requires that the activity phase started long before the estimated age
(10-100 yr) for the component farthest from the nucleus.  In fact, a
plausible time to heat the gas is at least of the order of the sound
crossing time $t_s$, and $t_{s}\sim 4\times 10^5$ yr to cross the hot
center out to $R=2^{\prime\prime}$; $t_s\sim 10^6$ yr to reach
$R=4^{\prime\prime}=312$ pc.  Heating of the center requires then at
least $\sim 10^6$ yr.  An upper limit to the time elapsed since the
heating episode is given by the radiative gas cooling time for the
center, that is $t_{cool}=3kT/2n\Lambda(T,Z)$, where $\Lambda(T,Z)$ is
the cooling function (taken here from Sazonov et al. 2005). For the
innermost $2^{\prime\prime}$, $t_{cool}\sim 10^7$ yr, and $t_{cool}$
is a factor of $\sim 3$ longer for the surrounding annulus.  Then a
radio outburst should have heated the gas not longer than $\sim 10^7$
yr ago. Indeed, in the hypothesis that this source is continuously
ejecting components from the core that soon disrupt and cannot reach
out of the nucleus, heating may be frequent, and confined to the
galactic central region, keeping the gas hot there.
The outburst duration in bright
radio sources is quite well constrained to be $10^7-10^8$ yr (e.g.,
Worrall 2009); the outburst recurrence time, for radio sources with
$L_{1.4GHz}>10^{25}$ W Hz$^{-1}$, is such that each source is
retriggered once every 0.5--few Gyr (Best et al. 2005, Worrall
2009). For lower luminosity sources, both the lifetimes and the
recurrence times are more uncertain; it seems that the activity must
be more frequently retriggered (Best et al. 2005). NGC4278 could then
be an extreme case of frequent triggering of low power jet
components\footnote{Note that a hot accretion flow (as considered in
  Sects.~\ref{nuc} and~\ref{nofeed}) may well have resumed after a
  major heating episode of $\lsim 10^7$ yr ago, since the flow time
  from $r_{acc}\sim 20$ pc to the center, at the free-fall velocity,
  is $\lsim 10^5$ yr.}.

The final question is whether a past radio outburst could have
injected enough heating to produce at least the thermal energy of the
central hot region currently observed, $E_{th}=4.1\times 10^{53}$ erg
(within $2^{\prime\prime}$, Tab.~\ref{phys}).  A jet power of the
order of that estimated here, $P_{jet}\sim (1-2)\times 10^{42}$ erg
s$^{-1}$, can account for $E_{th}$ in 0.4--1$\times 10^4$ yr, a time
much lower than typical lifetimes of individual radio sources (see
above); clearly, this requires that the whole $P_{jet}$ is transferred
to the ISM. $P_{jet}$ is also far larger than the gas luminosity
within $2^{\prime\prime}$ ($1.0\times 10^{39}$ erg s$^{-1}$) and
within $4^{\prime\prime}$ ($1.16\times 10^{39}$ erg s$^{-1}$).  

\subsubsection{Hot nuclei in other early type galaxies}

In four other spheroids $Chandra$ observations revealed a clear
temperature increase towards the center. All these cases reside in low
power radio galaxies ($L_{radio}<2\times 10^{38}$ erg s$^{-1}$), or
where the radio emission is not detected at all. The first one was
Sombrero (NGC4594, Pellegrini et al. 2003), where the LINER nucleus of
$L(2-10$ keV)$=1.5\times 10^{40}$ erg s$^{-1}$ is embedded in 0.7 keV
gas, out to $r\sim 160$ pc, and then sorrounded by $kT\lsim
0.4$ keV gas. This nucleus also hosts a compact radio source of low
luminosity ($L_{15GHz}=1.58\times 10^{38}$ ergs s$^{-1}$).  An
increase in the central temperature was found in NGC 4649 (Humphrey et
al. 2008), peaking at $\sim 1.1$ keV within the innermost 200 pc, and
dropping to $\sim 0.8$ keV outside; NGC4649 is a faint radio source
extending for $\sim 4$ kpc ($L_R=1.4\times 10^{37}$ erg s$^{-1}$, Dunn
et al. 2010).  The nuclear spectrum of the Virgo elliptical NGC4552
includes a power law from the AGN plus a thermal plasma with $kT =
1.04$ keV; 5GHz VLBA observations reveal a two-sided extended
emission,  suggestive of a pc-scale jetlike
structure, of total radio luminosity of $1.55\times 10^{38}$ ergs
s$^{-1}$ (Machacek et al. 2006).  Finally, a projected temperature of
0.65 keV within $2^{\prime\prime}$ and 0.40 keV within
$2^{\prime\prime}-4^{\prime\prime}$ (for d=9.7 Mpc) was found in
NGC3115 (Wong et al. 2011), with $L_X<10^{38}$ erg s$^{-1}$ and no
detected radio emission; the central rise in temperature was
attributed to the accretion flow.

A central temperature sigificantly larger than in the surroundings may
be then a typical feature appearing during the activity cycle commonly
followed by early type galaxies in the local universe.  It could be
the sign that sometimes, in the lifecycle of low power objects, the
jet cannot bore out of the nucleus' surroundings, but it can heat
them.

\subsection{The extended, misaligned hot gas and the colder phases}\label{cold}

One of the new and interesting findings from the deep observation of
NGC4278 is the presence of hot gas, with a spatial distribution more
elongated than the optical one, and also misaligned with it (Figs. 1
and 4).  The soft diffuse X-ray emission seems instead aligned with
that of the ionized gas, residing in a rotating disk or spiral, of
semi-major axis length of $\sim 25\arcsec$ ($\sim 2$ kpc;
Fig. 4; Sarzi et al. 2006; see also Osterbrock 1960, Goudfrooij et
al. 1994).  Thus there is a spatial association (in projection)
between the hot and the ionized gas phases: both show an elongated
distribution, following a direction described by the same position
angle, and misaligned with respect to the optical major axis.  The
SAURON data show that the ionized gas velocity field is regular,
consistent with circular gas motions in a low inclination disk, with a
deviation from this simple situation in the outer measured region, due
to a gradual twisting of the velocity field.  The gas is
rotating in the same sense as the stars, but with a kinematic
misalignment, of a median angle of 29$\pm 19$ degrees within
$22^{\prime\prime}$; the gaseous and stellar kinematics are misaligned
by increasingly wider angles, towards the outer parts of the field
(Sarzi et al. 2006).  The ionized gas is not rotating in the
equatorial plane of the galaxy, and its distribution and kinematics
are likely tracing gas that is accumulating while flowing in (Sarzi et
al. 2006). Finally, note that the SAURON angular resolution 
(each pixel is $\sim 1\arcsec$ wide) was adequate to establish that the ionized gas is
diffuse and not filamentary (see also Goudfrooij et al. 1994).

NGC4278 also contains a large HI mass ($6.9\times 10^8 M_{\odot}$,
Sect. 1), residing in a large regular disk or ring (Raimond et
al. 1981), extending for 37 kpc, which implies that the system is old
and evolved; the tails from the disc, revealed by a deep map, indicate
that the system is still accreting gas (Morganti et
al. 2006). Notwithstanding the extensive reservoir of neutral
hydrogen, and the very regular disc kinematics suggesting that the
galaxy has been surrounded by cold gas for a few Gyr, there is no
evidence for the presence of a young stellar population.  Studies from
various groups report a uniformely old age, from scales of $R_e/16$ to
the whole galaxy (Terlevich \& Forbes 2002,  Serra et al. 2008, 
Jeong et al. 2009, Kuntschner et al. 2010, Shapiro et al. 2010).  Therefore, rather
than recent merging with starformation, smooth accretion from outside
is suggested, with the gas spread over a large area, diluted and
resulting in the formation of little CO emission and hence no
star-formation (Morganti et al. 2006; CO emission is not detected, see
Sect. 1, Crocker et al. 2011).

The HI disk extends down to the innermost resolution element of the
Westerbork observation ($\approx 2\times 1$ kpc), where the ionized
gas resides; in the more central regions, the HI and the ionized gas
are suggested to be physically associated, since they are spatially
and kinematically consistent (Oosterloo et al. 2010, Morganti et
al. 2006, Goudfrooij et al. 1994). Given this, and the fact that the
radial profile of the H$\beta$ flux can be fully explained by
photoionization by the LINER nucleus only within the central
$3\arcsec$, but is clearly more extended than that, the ionised phase
could be produced by shock-heating of the accreting gas (Sarzi et
al. 2010; see also Eracleous et al. 2010b).  A weak triaxial
perturbation of the stellar potential could funnel the gas into
preferential streams, where fast shocks between gas clouds may
occur. A problem with this origin for the warm gas is that shock
velocities as high as required to power the observed line emission are
not expected in the potential of NGC4278; other diffuse sources of
ionization, as a hot ISM or old post-AGB stars, were then suggested to
contribute or be dominant (Sarzi et al. 2010).

Finally, NGC4278 shows a complex and irregular dust structure in its
core, with several dense knots, interconnected by filaments, and
prominent patchy dust out to $\sim 25\arcsec$, mostly on the northern
side of the nucleus; closer to the galactic center, due to a spiral
pattern, the dust appears to be streaming toward the nucleus
(Goudfrooij et al. 1994, Carollo et
al. 1997, Lauer et al. 2005). $Spitzer$ images revealed extended
emission from warm dust, molecular hydrogen and ionized gas (see
Fig. 4b), and IRS spectral observations shows excitation and ionization in
the mid-IR (Tang et al. 2011).

Ionised gas, cold gas and dust then seem different phases of the same
accretion process of gas from outside; moreover, the diffuse X-ray
emission follows the general pattern of the [OIII] and 8$\mu m$
emission, and thus is likely to be associated with the bar-like
distribution of the other phases (Fig. 4). Is the hot gas interacting
with these phases, and does its elongation find an explanation in this
interaction?  In fact, the flattening of the diffuse X-ray emission
with respect to the optical surface brightness, and even more its
misalignement, are an unexpected result: the hot gas originates in the
stars, and at equilibrium its density should follow the isopotential
surfaces, that are rounder than the optical isophotes, for reasonable
galaxy models (e.g., Binney \& Tremaine 1987).  The origin of the hot
gas elongation is not likely to reside in the presence of significant
rotational support, inherited from the stars, from which the hot gas
comes: in NGC4278 only the inner galactic region is occupied by a fast
stellar rotator, with a maximum rotation of $\sim 70$ km s$^{-1}$ at
0.4$R_e$, and steeply decreasing outside.  Also, the stellar
kinematical axis is misaligned with respect to an axis perpendicular
to the hot gas elongation (Fig. 4).  Another potential contributing
factor for the hot gas elongation could be an AGN outflow, if it
cleared the ISM along directions perpendicular to the observed hot
(and warm) gas elongation, in a sort of bipolar outflow (e.g., Novak
et al. 2011).  Since the ionized gas disk has a modest inclination,
such an AGN outflow, while not ruled out, seems less likely; also, no
significant kinematic disturbance is seen in the [OIII] SAURON
velocity field, and then the elongated X-ray emitting region seems
more a zone of enhanced emission.  We also note that the same
correlation between the X-ray emission, ionized gas, and 8$\mu m$
emission in a flattened structure (Fig. 4) may be indicative of a
dynamically supported ISM substructure (i.e., a gaseous bar, or warp,
or some combination) located within the inner region of NGC4278. Even
in this model the X-ray feature is not directly associated with the
directionality of the AGN jets.

The energy flux from the hot to the cold gas, by increasing the X-ray
emissivity, could be the reason why the hot gas seems not co-spatial
with the stellar distribution, but instead well aligned with the warm
gas. The cold gas, accreting more or less steadily for a few Gyr,
created within the galaxy a disk (or bar-like) region where the hot gas is made to
cool preferentially, thus showing more emission there (e.g., Fabian et
al. 2003).  
Thermal conduction is a specific example of an energy flux mechanism,
and it has often been suggested to cause energy to flow from the hot
coronal gas into colder gas, in elliptical galaxies and galaxy
clusters (Sparks et al. 1989, Macchetto et al. 1996, Fabian et
al. 2003).  The ionized emission could result then from the cold
accreting material that is being excited by the hot gas, during the
evaporation process expected from the interaction between cold and hot
gas. Thermal electron conduction excites the cold gas into optical
emission, and locally enhances the X-ray emissivity and cools the hot
gas; for example, two temperatures are present in the X-ray gas of
M87, with the lower one in the vicinity of optical filaments (Sparks
et al. 2004). Thus, electron conduction could explain, at least in
part, also the discrepancy between the model and the observed
temperatures out of few hundreds pc.  We examine next the energetics
involved more quantitatively.

Under the assumption that the energy flux is in the saturated regime
(Cowie \& McKee 1977), the energy available to excite the optical
emission is $Q_{sat}=5.2\times 10^{40} T_7^{3/2} n_{0.01}
D_{maj}D_{min}$, where $T_7=T/10^7$K is the hot gas temperature,
$n_{0.01}=n_e/0.01$cm$^{-3}$ is its density, and $D_{maj}D_{min}$ is
proportional to the apparent projected surface area ($D_{maj}$ and
$D_{min}$ being the major and minor diameter of the line emitting
region in kpc; Macchetto et al. 1996).  The H$\alpha$ emission line
flux that is produced is $L\sim 10^{39} T_7^{3/2}
n_{0.01}D_{maj}D_{min}$, if 1\% of $Q_{sat}$ is radiated in the
H$\alpha$ line, and the total surface area is twice the apparent
projected surface area (this is a conservative assumption, that
applies to the observed ellipse shown in projection by the disk, whose
depth along the line of sight, and thickness, remain unknown). The gas
temperature and density outside the central $3^{\prime\prime}$ radius,
where the LINER has been shown to fully account for the ionization
(Sarzi et al. 2010, their Fig. 4), are $T_7=0.37$ (for $kT=0.32$ keV),
and $n_{0.01}=1$; for $D_{maj}\sim 4$ kpc and $D_{min}\sim 2$kpc
(Fig. 4), one has $L\sim 10^{39}$ erg s$^{-1}$. The total H$\alpha$
luminosity observed is $9.1\times 10^{39}$ erg s$^{-1}$ (rescaled for our
adopted distance), and $5.8\times 10^{39}$ erg s$^{-1}$ of this comes
from outside the central $3^{\prime\prime}$ radius (from data kindly
provided by M. Sarzi).  Thus the energy flux does not violate the
limit given by the observed line emission, while contributing to the
ionization of the extended warm gas distribution.

The presence of thermal conduction could also explain why
the hot phase is colder than simple models predict, as shown by 
Fig.~\ref{hyd} (Sect. 5.2.1). The total energy flux
from the hot gas to the cold one, $Q_{sat}$, turns out to be 
$\sim 10^{41}$ erg s$^{-1}$, a non-negligile amount. For
example, the SNIa's heating for the whole galaxy amounts to $10^{41}$
erg s$^{-1}$ at an age of 10 Gyr. Indeed, remarkably, the hot gas
temperature seems close to just the stellar heating prediction
(Fig.~\ref{hyd}), without traces of additional (e.g., from
SNIa's)  heat sources.

The timescale for this process of heat transfer from hot to cold gas
(or the evaporation lifetime) is difficult to compute, because of the
unknown internal density and distribution of the HI gas; however, given that the HI is
much more massive than the hot gas, and still accreting, the process
is not a fast, recent, and transient phenomenon (recall that also the
hot gas is continuously replenished by stellar mass losses).

\section{Conclusions}\label{ccc}

We analyzed a deep (579 ks) $Chandra$ ACIS pointing of the radio
compact elliptical galaxy NGC4278, hosting a two-sided pc-scale
jet. We detected soft X-ray emission from hot gas out to a radius of
$\sim 5$ kpc, with a 0.5--8 keV luminosity of $2.4\times 10^{39}$ erg
s$^{-1}$, elongated and misaligned with respet to the stellar
body. At the nucleus, AGN-like emission with a spectral shape
consistent with that of previous pointings since 2004 ($\Gamma
=2.31\pm 0.20$, and small intrinsic absorption) decreased by a factor
of $\sim 18$ since the first $Chandra$ observation. Hot gas is also
present at the nucleus, with a temperature ($kT=0.75$ keV)
significantly larger than in the surrounding region, where $kT\sim
0.3$ keV.  We investigated the relationships between nuclear emission,
fuel availability, and mechanical energy output from accretion, as
well as the interaction between hot and cold gas phases, with the
following main results.

The low luminosity of the AGN [$L(0.5-8$ keV)$/L_{Edd}=5.9\times
10^{-8}$ in 2010] can be explained by a low radiative efficiency
accretion flow, with mass accreting at a rate close to that of the
Bondi accretion solution [$\dot m_B=(0.7-7)\times 10^{-4}$]. An
average $\dot M_{B}=2.1\times 10^{-3}\, M_{\odot}$ yr$^{-1}$
corresponds to $L_{0.5-8 keV}\lsim 5\times 10^{40}$ erg s$^{-1}$,
while the observed $L_{0.5-8 keV}= 2.5\times 10^{39}$ erg s$^{-1}$;
however, the effect of rotation in the gas close to the accretion
radius can easily reduce this expected $L_{0.5-8 keV}$ to $\lsim 5\times
10^{39}$ erg s$^{-1}$.  Estimates of the jet power from scalings with
the radio luminosity at 5 GHz and 200--400 MHz, and with $\dot M_B$,
all agree giving $P_{jet}=(1-2)\times 10^{42}$ erg s$^{-1}$.  This
power requires that the rate of accretion $\dot M$ on the MBH 
is not largely reduced with respect to $\dot M_B$.

The hot gas flow, followed with hydrodynamical simulations including
heating by SNIa's and only gravitational effects from the MBH, is an
outflow on the galactic scale during the entire evolution, consistent
with the low observed hot gas content.  There is also a small
accretion region at the center, through which mass flows on the MBH at
a rate consistent with the estimate of $\dot M_B$, at the present epoch. The MBH heats the
surrounding gas on a much smaller region than observed, and the
large and sharp temperature increase at the center is not reproduced; also, out of the central
$\sim 0.5$ kpc the model temperature is larger than observed,
indicating that the SNIa's heating is less important than standardly
assumed, or that the gas is cooling due to the interaction with other
gas phases. An MBH outburst in the recent past due to accretion at
high $\dot m$ would be supported by the finding of a central
overpressured region in the hot gas, and of a central ``hole'' in the
cooler gas phases; however, it is not likely to be the origin of the
central large temperature, since it should have also left a flatter
brightness profile than observed in the central galactic region, and
traces of starformation.

A jet could have heated the central gas, during a past activity
episode, taking place a time $\lsim t_{cool}\sim 10^7$ yr ago. Extended radio
emission is not detected, and the radio spectrum does not show signs
of older lobes, thus it is likely that also in the past radio
outbursts have not succeeded in getting out of the central galactic
region. The central temperature may then have originated by a
frequent triggering of low power jet components, that remain confined
within the nucleus. A high temperature at the center may be a typical feature appearing
during the activity cycle of low power objects, when the jet cannot
bore out of the nucleus surroundings, but it can heat them. Unless
$P_{jet}$ has been overestimated, the thermal confinement of
the pc-scale jets requires the hot gas pressure to increase by a large
factor ($\gsim 10^3$) above its average over the central 156 pc.
Other colder gas phases seem to have been cleared from the nuclear
region.

On the galactic scale, there is a spatial association (in projection)
between the hot, the cold and the ionized gas phases, both elongated in
the same direction, misaligned with respect to the optical major axis.
The accreting cold gas triggering the cooling of the hot
phase could produce both the unusual elongation and misalignement of
the X-ray emission, and the lower temperature of the hot gas with
respect to the predictions of hydrodynamical simulations.  Part of the
ionized emission could result from the cold gas excited by the
hot gas, during the evaporation process.

Investigations of the interplay between the hot gas and the nuclear
activity are fundamental in the context of understanding the MBH-host
galaxy coevolution; NGC4278 is the first case of a low/medium mass
galaxy, with a modest hot gas content, that has been explored in
detail using $Chandra$ observations.  Even though expensive in terms
of exposure time, other studies of similar objects are very much
needed to build a coherent and comprehensive picture of accretion in
the local universe. The interplay between different gas phases, as
resulting from multiwavelength observations, is another field rich of
consequences for our knowledge of the galactic ecology and evolution,
and allowed here for a more exhaustive explanation of the hot gas
properties as well.

\acknowledgments We thank L. Ciotti, G. Giovannini and R. Sancisi for
useful discussions, and M. Sarzi for kindly providing data and the
ionized gas images used for Fig. 4.
This work was partially supported by the Chandra GO grant GO0-11102X
(PI: Fabbiano) and NASA contract NAS8-39073 (CXC).
GT and SP also acknowledge partial financial support from the ASI-INAF grant I/009/10/0.
The data analysis was supported by the CXC CIAO software
and CALDB. We have used the NASA NED and ADS facilities,
and have extracted archival data from the Chandra archives.
{\it Facility}: CXO (ACIS).

\begin{figure*}
\vskip 1truecm
\centerline{\includegraphics[scale=0.5]{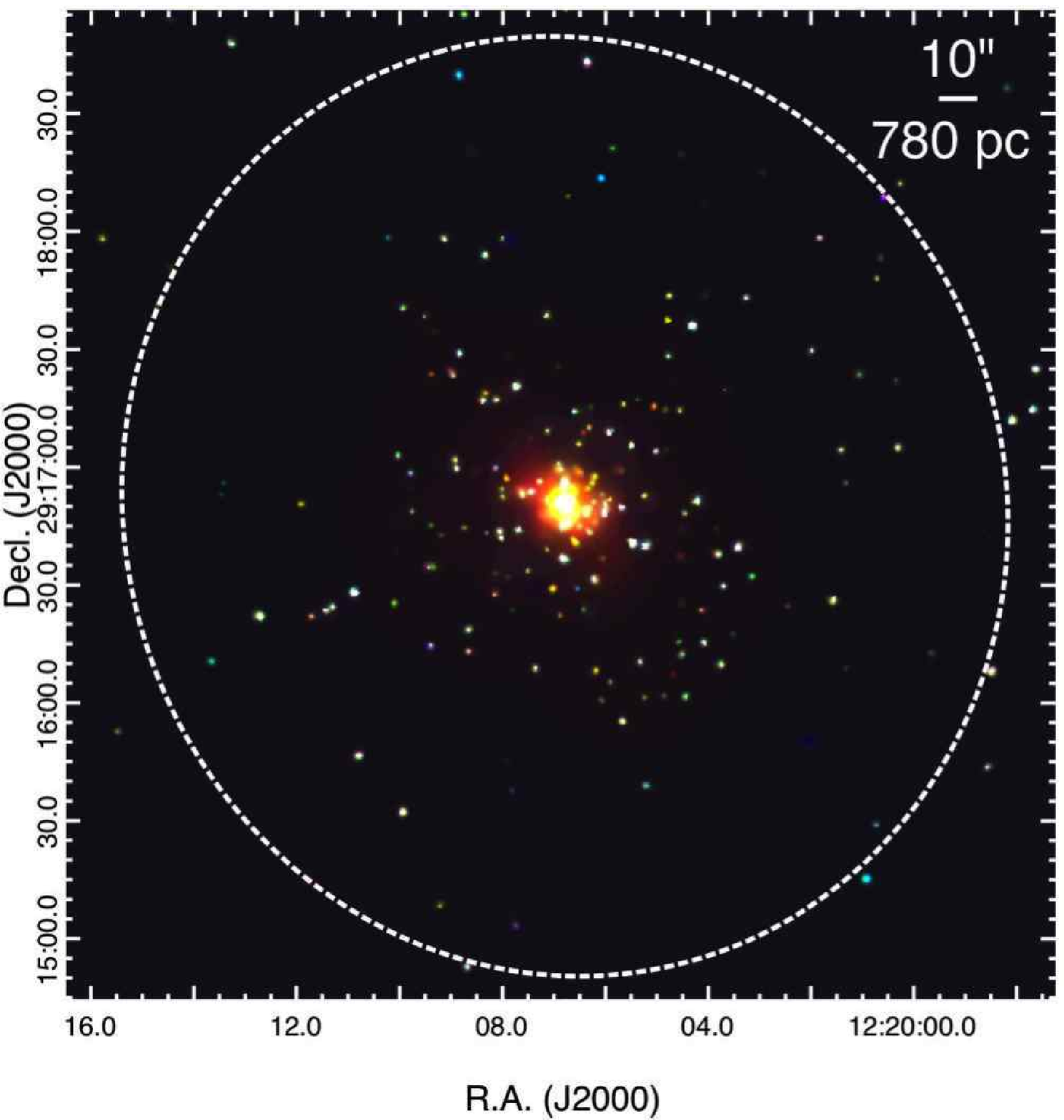}}
\centerline{\includegraphics[scale=0.45]{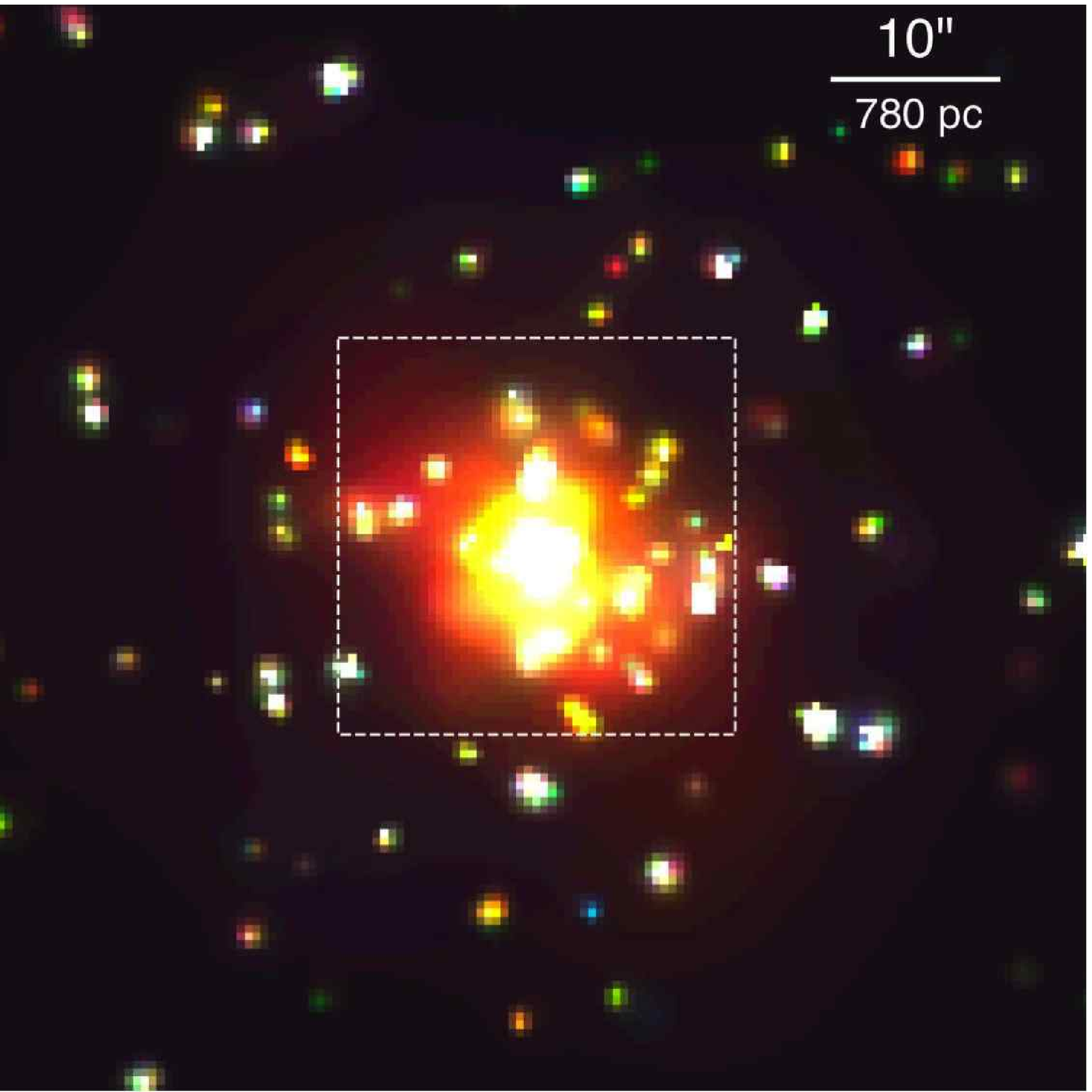}}
\caption{Top (a): three-color composite image ($4\arcmin \times 4\arcmin$)
  of NGC 4278. Red represents the 0.3-0.9
  keV X-ray emission, green represents the 0.9--2.5 keV emission, and
  blue the 2.5--8 keV emission. North is up, East is to the left. 
The white dashed ellipse indicates the
  extent and orientation of the optical D25 ellipse. Bottom (b): the zoomed-in X-ray image of the
  central $70\arcsec$ region; the dashed square shows the region zoomed
  in the next Fig. 1c.  See Sect. 2 and 3.1 for more details.}
\label{f1}
\end{figure*}

\addtocounter{figure}{-1}
\begin{figure}
%\plotone{f1c.eps} 
\centerline{ \includegraphics[scale=0.5]{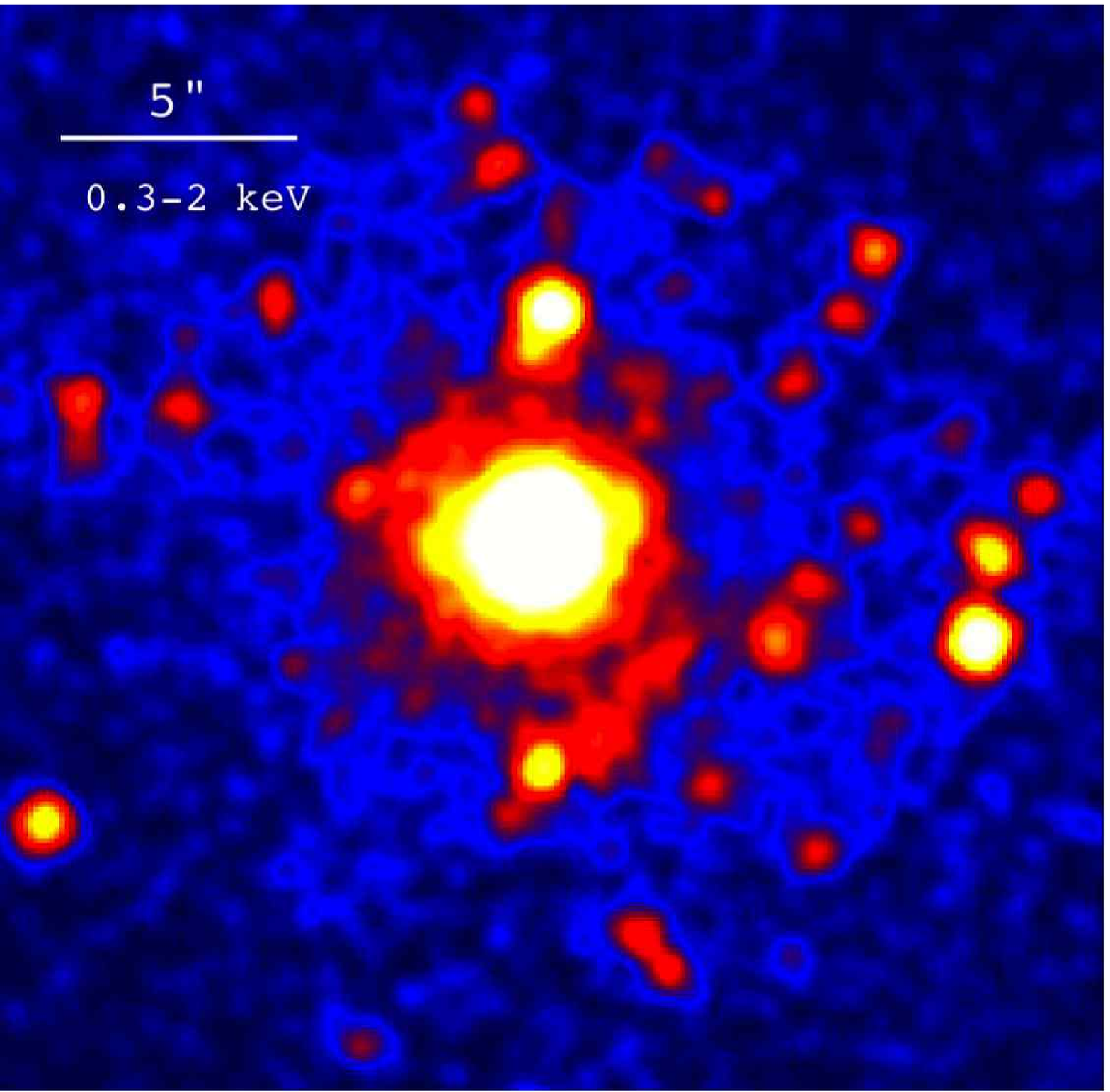} } 
\caption{Continued. (c) The innermost $20\arcsec$ region showing the
  subpixel rebinned X-ray emission (0.3--2 keV; smoothed with a
  $FWHM=0.3\arcsec$ gaussian kernel). See Sect. 3.1 for more details.}
%\label{f1c}
\end{figure}

\begin{figure}
\vskip -1.5truecm
\centerline{ \includegraphics[scale=0.5]{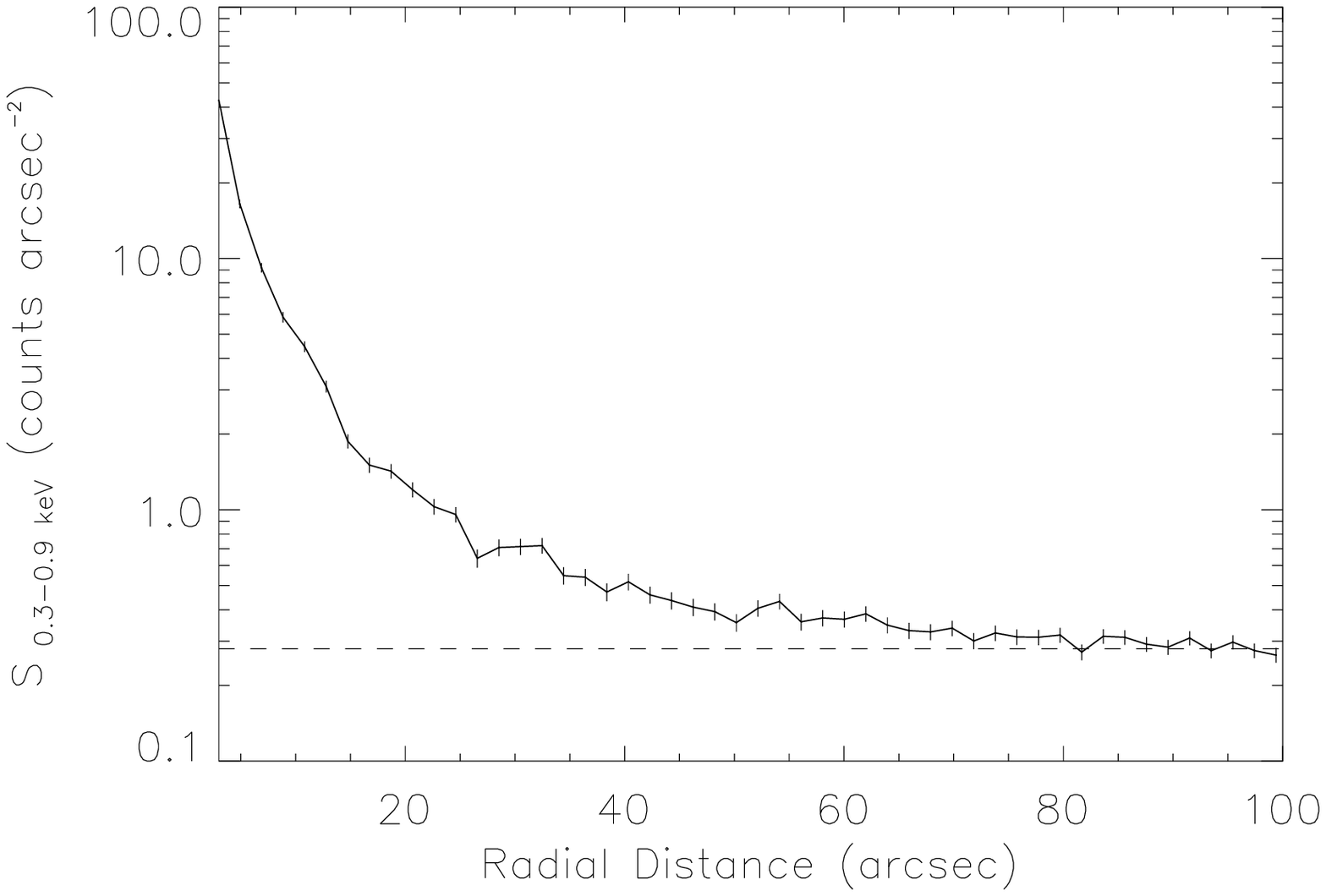} } 
\centerline{  \includegraphics[scale=0.5]{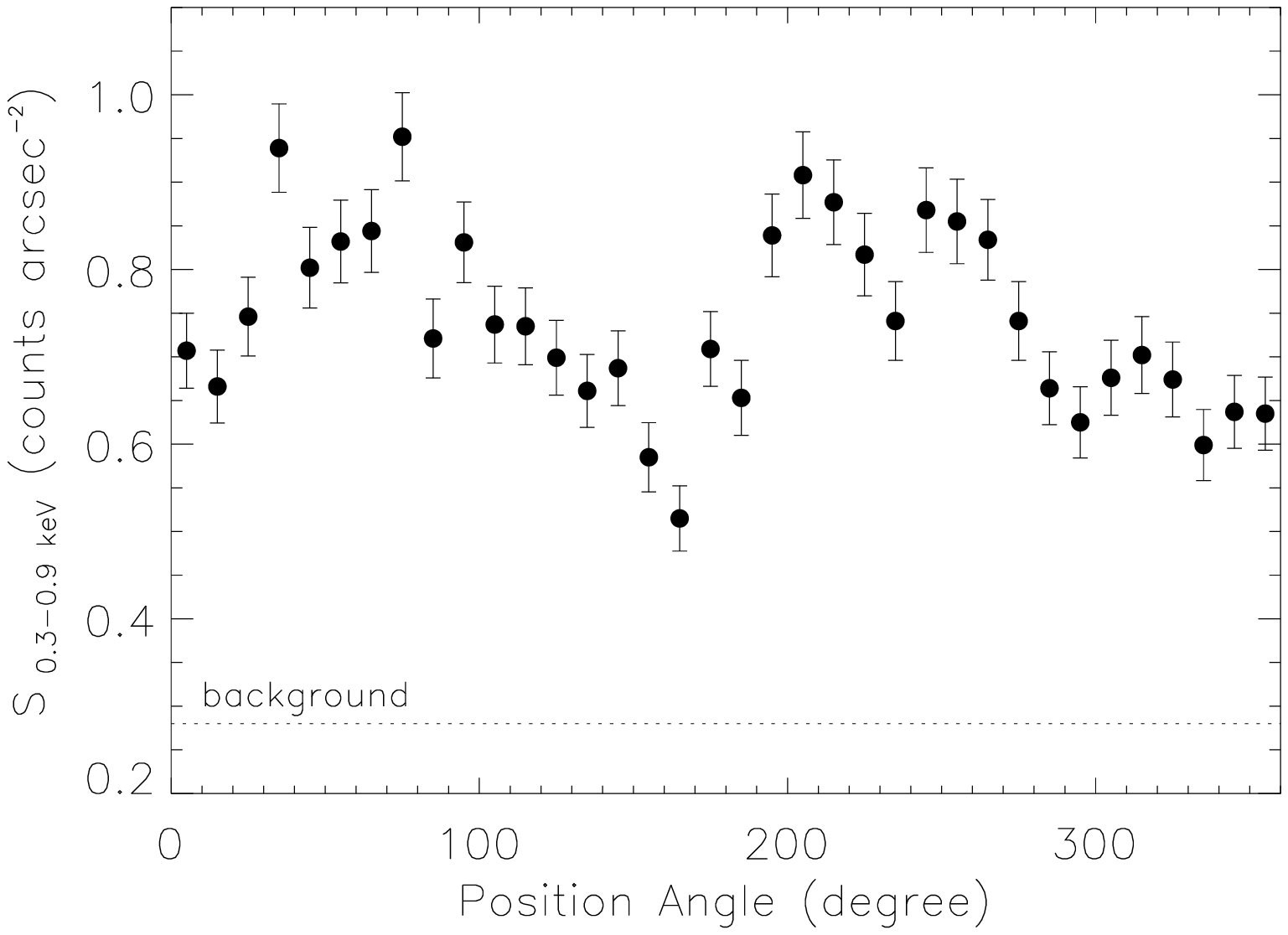} } 
\centerline{  \includegraphics[scale=0.5]{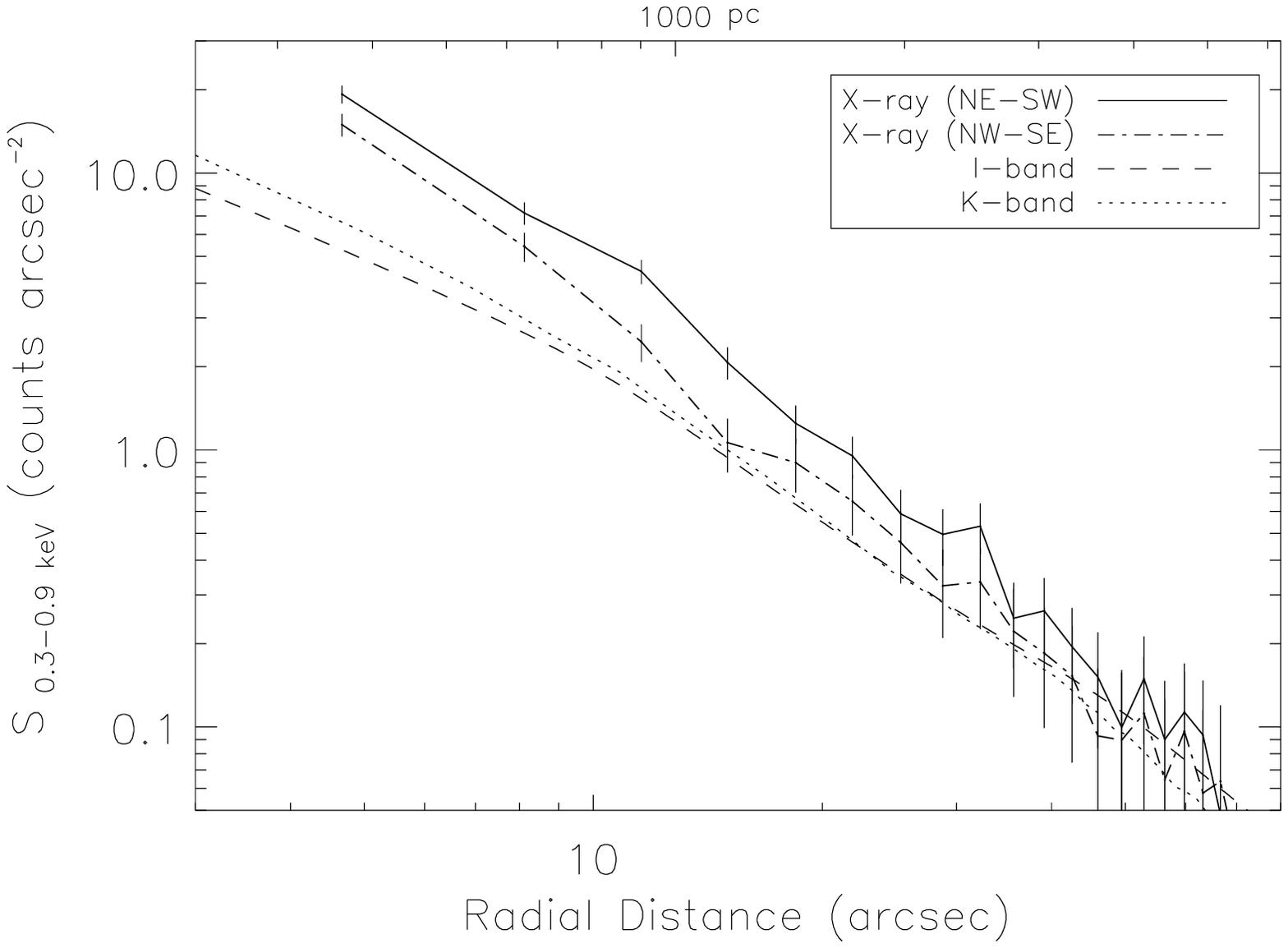} } 
\caption{Top: (a)  azimuthally averaged radial profile of the diffuse X-ray emission
  (point sources masked) in the 0.3--0.9 keV band. Middle (b):
average surface brightness for the extended soft X-ray emission,
measured in 36 ``fan-shape'' sectors, each $10^{\circ}$ wide, starting
from N and moving towards E,
and each one extending from a radius of $2^{\prime\prime}$ (thus
excluding the nucleus) out to $60^{\prime\prime}$.
Bottom (c): background subtracted, azimuthally averaged radial
  profiles in the NE-SW and the NW-SE sectors. The dashed and dotted
  lines are the radial profiles in the optical I-band (Cappellari et
  al. 2006) and the near-IR K band (Jarrett et al. 2003). See Sect. 3.1 for more details.}
%\label{brig}
\end{figure}

\begin{figure}
\centerline{ \includegraphics[scale=0.6]{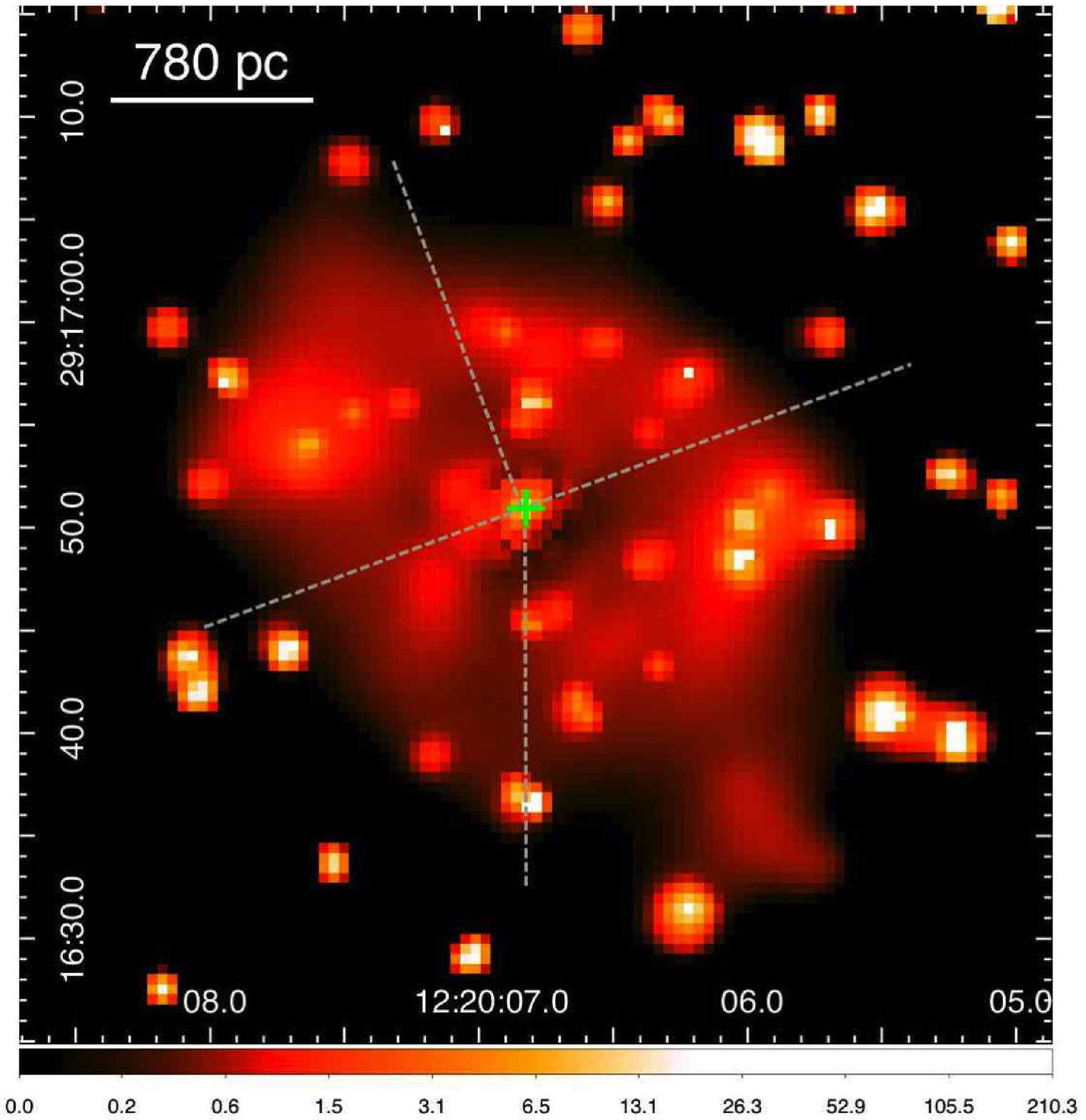} }
\caption{The intensity ratio image between the background
  subtracted soft X-ray emission and the 2-D $beta$ model image. 
The soft X-ray image was adaptively
smoothed with a minimum significance of 2.5$\sigma$. The four grey
lines indicate the boundaries of the quadrants defined in Section~\ref{morph}.
The optical center is shown with the green cross, and the 780 pc bar (top
left) is equivalent to $10\arcsec$. See Sect. 3.1 for more details.}
%\label{}
\end{figure}

\begin{figure}
\centerline{ \includegraphics[scale=0.4]{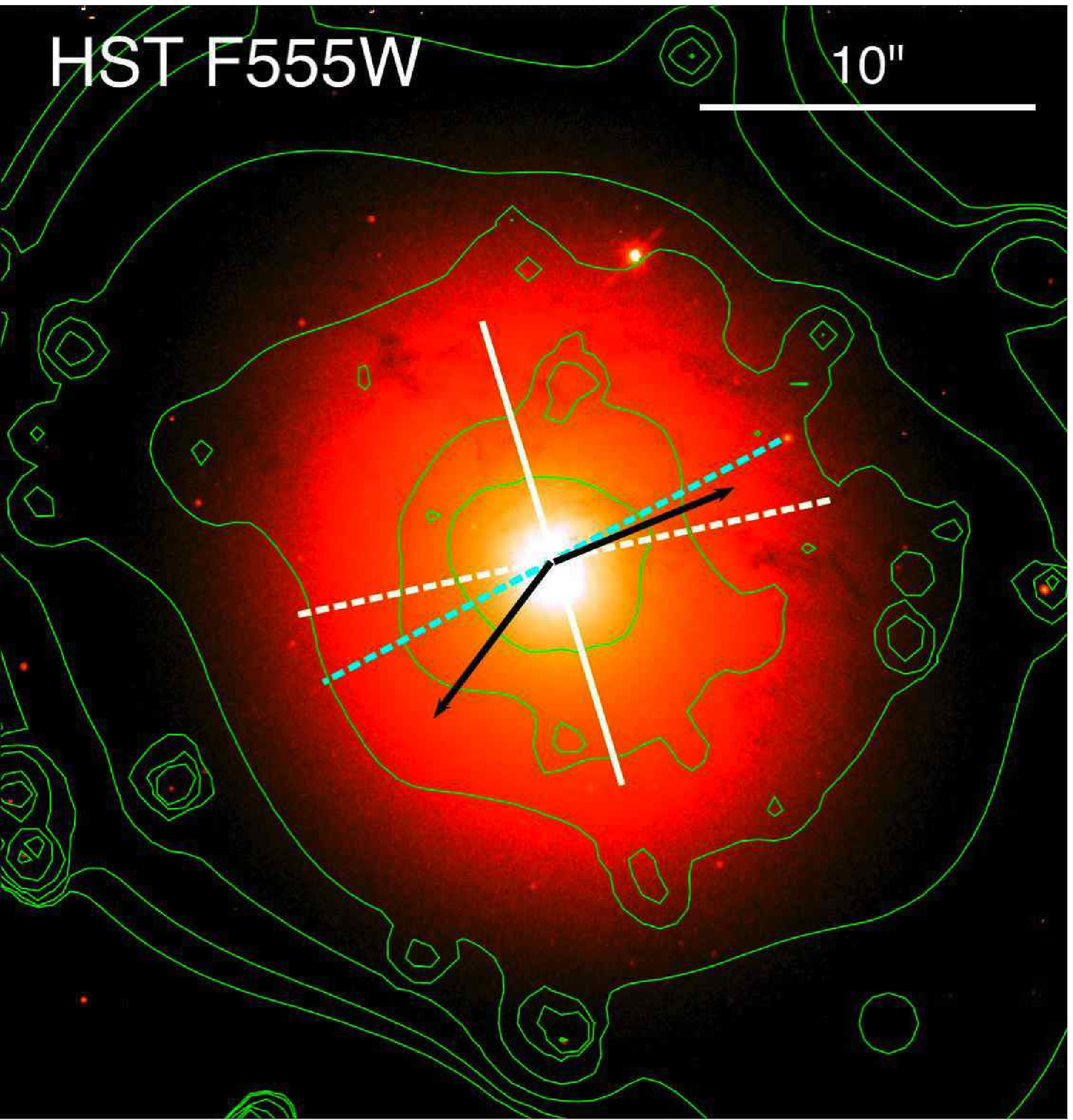}}
\centerline{\includegraphics[scale=0.4]{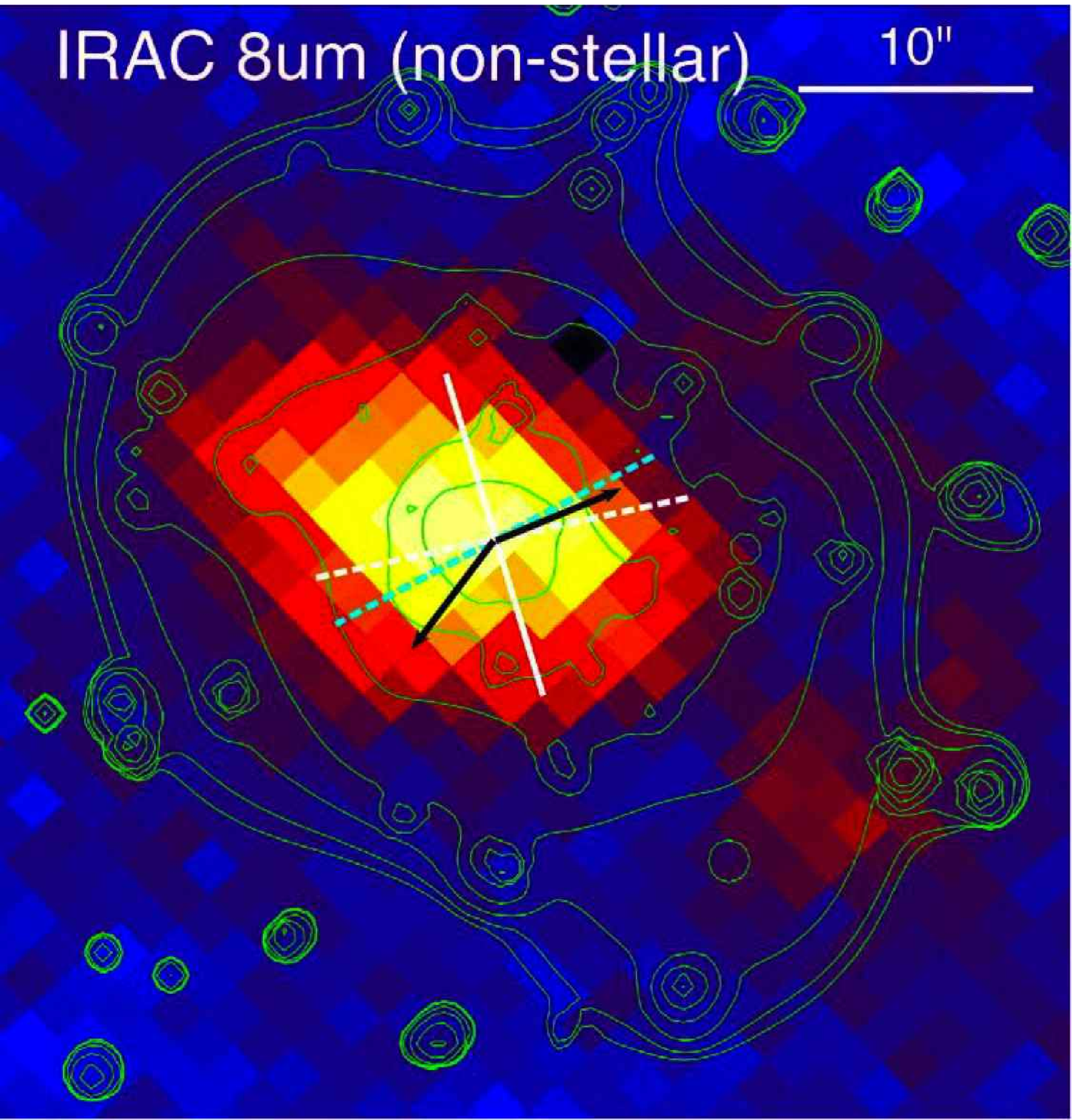}}
\caption{Contours of the extended X-ray emission in the 0.3--0.9
  keV band overlaid on: top (a) the $HST$ WFPC2 V band image (Carollo et
  al. 1997); bottom (b) the $Spitzer$ IRAC 8$\mu$m emission (with stellar continuum
  removed; see Tang et al. 2011).  Note that the $HST$ image has a smaller
  field of view. Relevant position angles are shown: the current radio
  jet directions (black arrows, from the knots S1 and N2 in the high resolution
radio map of Giroletti et al. 2005); the major axis of the optical light
  distribution in the inner galactic region (PA$_{phot}$=16.7$^{\rm o}$,
Cappellari et al. 2007;  white solid line), and the rotation axis of the stars in
  the inner galactic region (a line perpendicular to the
kinematic PA$_{kin}$=12.0$^{\rm o}$, Cappellari et al. 2007; white
dashed line); the rotation axis for the ionized gas
  (Sarzi et al. 2006; cyan dashed line). 
See Sect.~\ref{multi} for more details.
 } 
%\label{define}
\end{figure}

\addtocounter{figure}{-1}
\begin{figure}
\plottwo{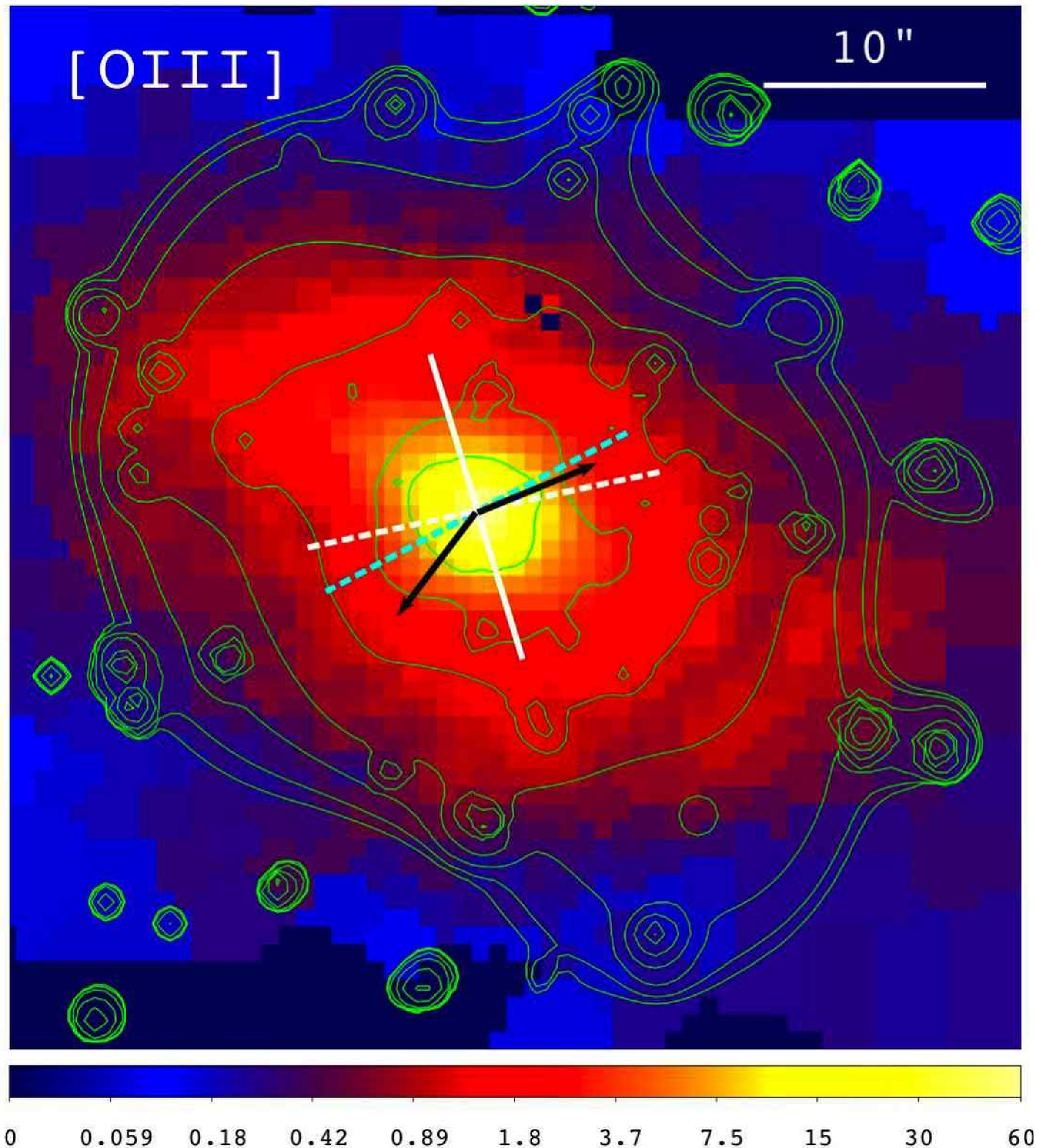}{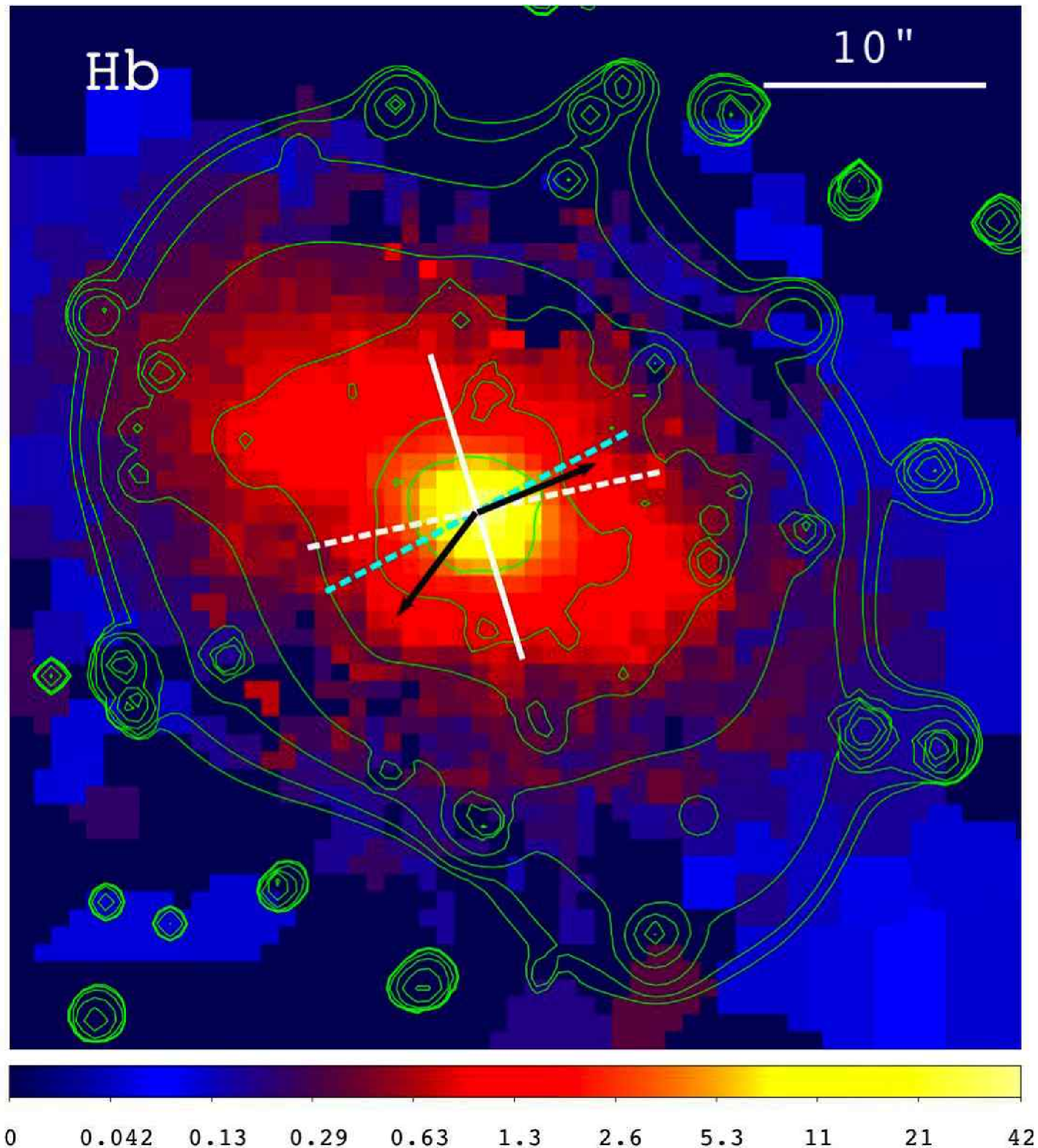} 
\plottwo{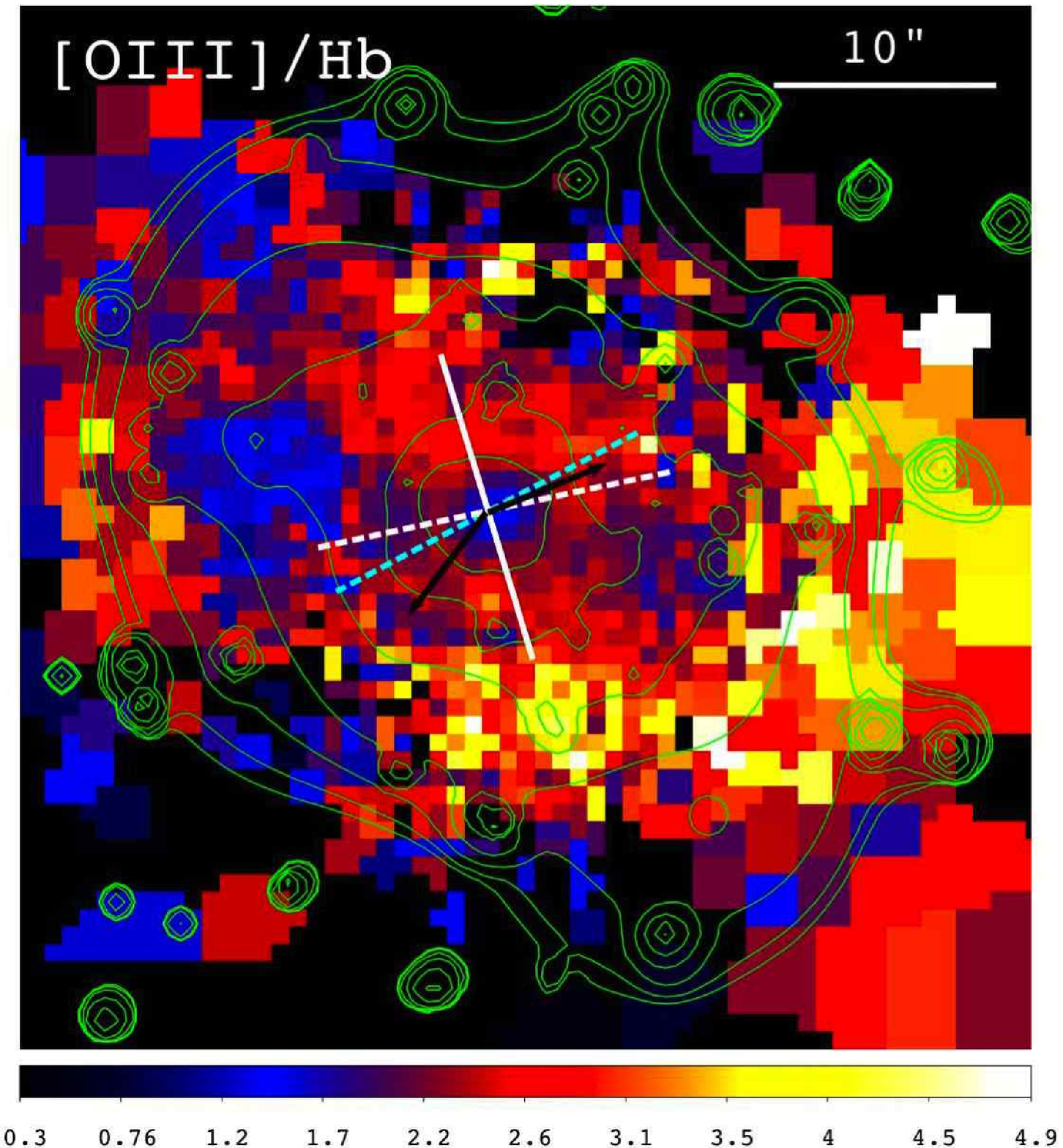}{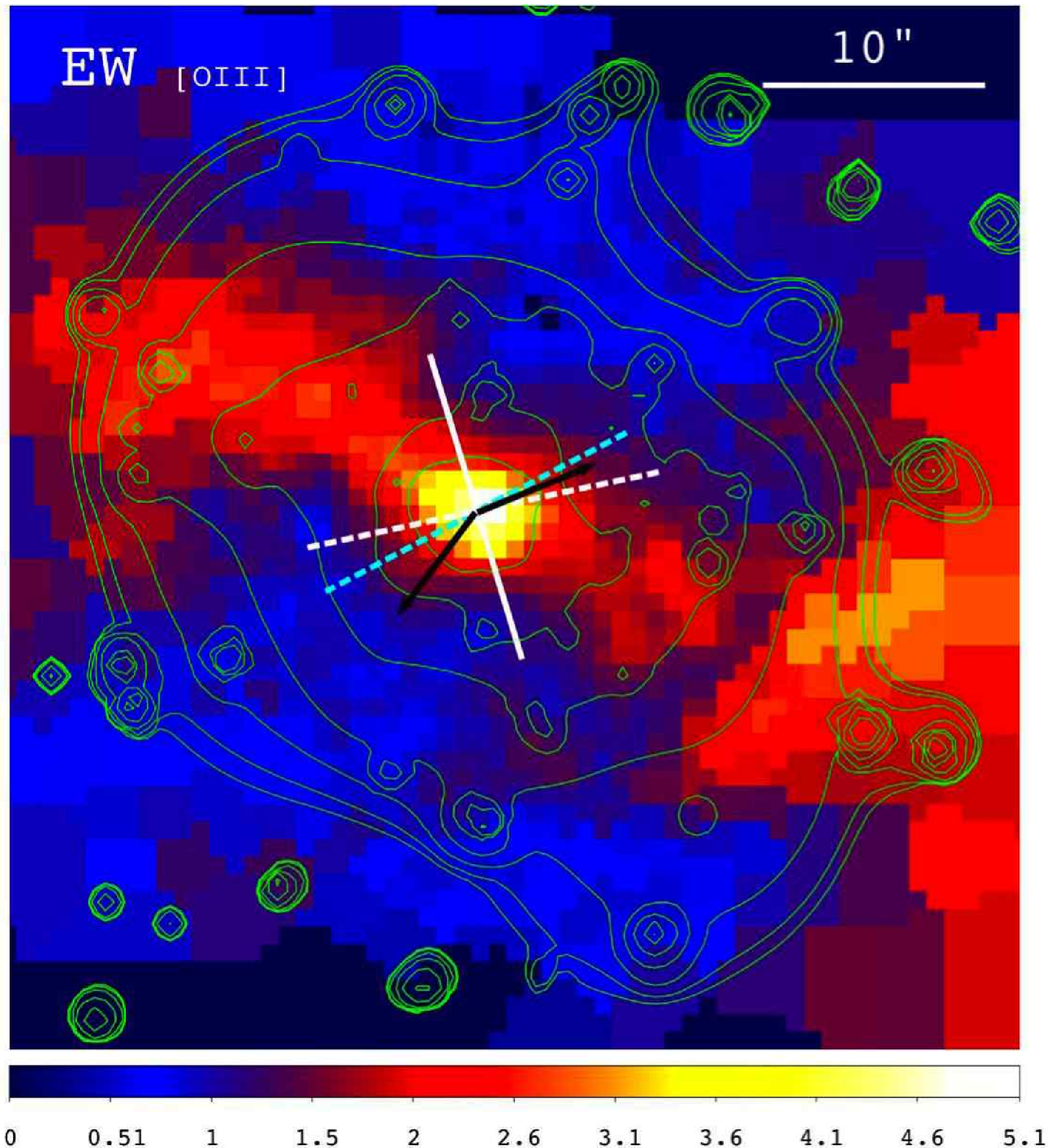}
  \caption{Continued. (c) Contours of the extended X-ray emission in
    the 0.3--0.9 keV band overlaid on the ionized gas maps 
    from the SAURON survey (Sarzi et al. 2006); top left: [OIII]
    flux; top right:
    H$\beta$ flux; bottom left: [OIII]/H$\beta$; bottom right:
    EW$_{[OIII]}$). The hot gas is elongated in a direction
 more aligned with the ionized gas distribution and the
$Spitzer$ IRAC 8$\mu$m emission than with
the optical stellar body of the galaxy. See Sect.~\ref{multi} for more details.
}
%\label{OB}
\end{figure}

\begin{figure}
%\vskip -3truecm
\centerline{ \includegraphics[scale=0.5,angle=-90]{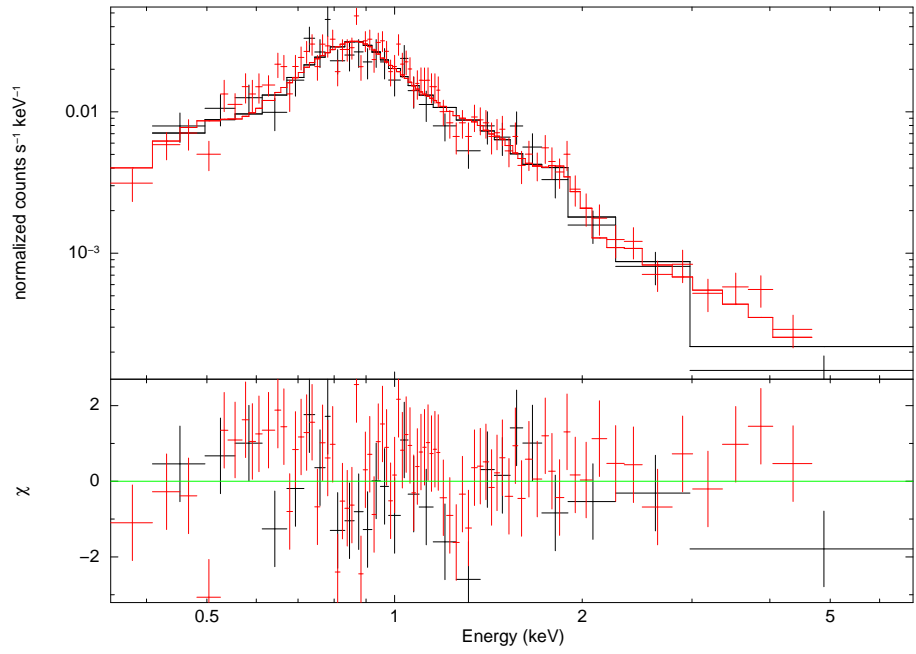} }
\caption{Best fit spectrum of the nuclear emission using the last
 two observations of March 2010 (Tab.~\ref{allflux})
that are not affected by pile-up. 
The red and black lines correspond to the nuclear spectra from ObsID
   11269 and 12124, respectively. See Sect. 4.1 for more details.}
%\label{fig:nuc_spec}
\end{figure}

\clearpage

\begin{figure}
%\vskip -5truecm
\centerline{ \includegraphics[scale=0.7]{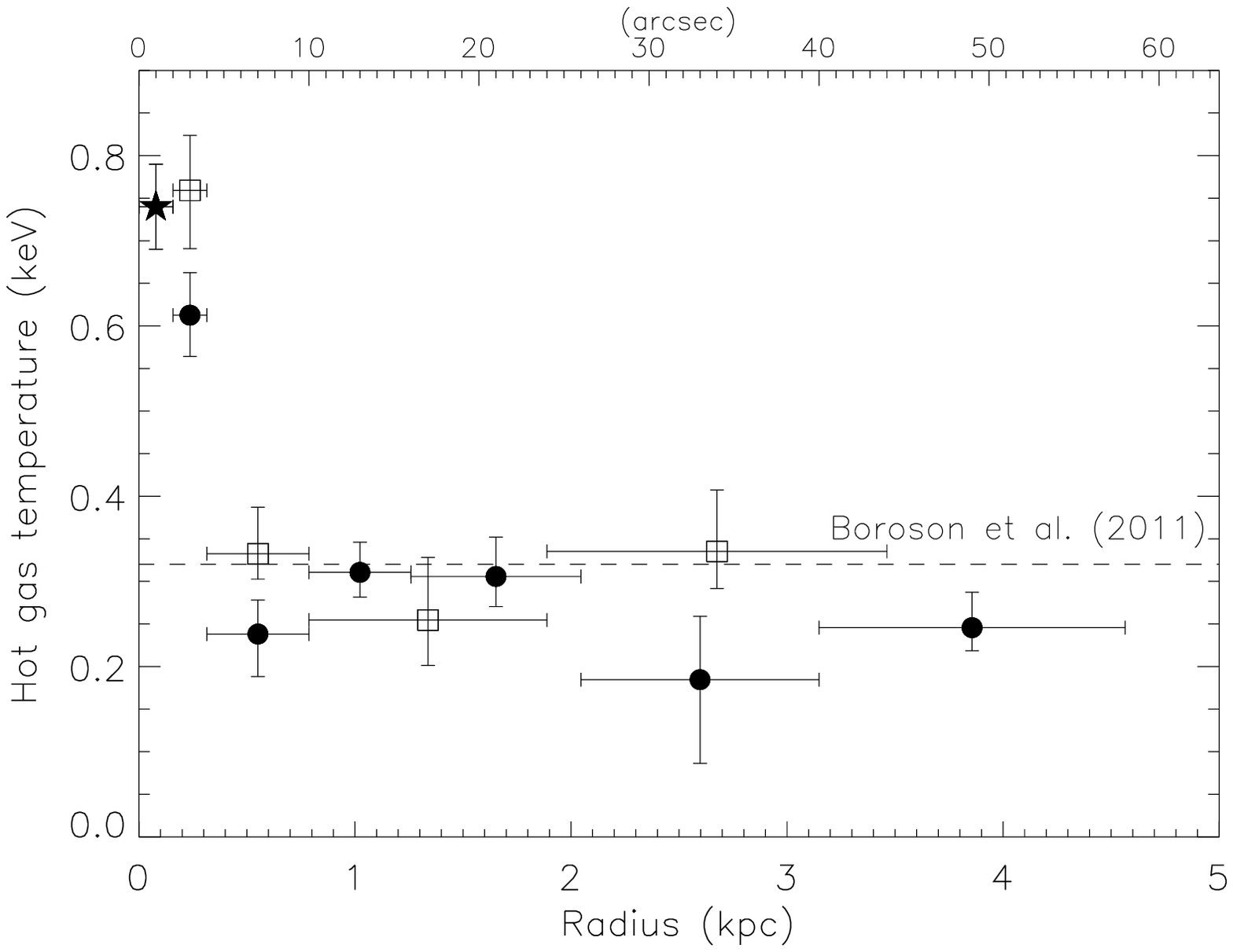} }
\caption{Radial distribution of gas temperature derived from
  spectral deprojection along two directions (NE--SW, filled circles; and
  NW--SE, open boxes). The temperature for the inner 2\arcsec\/ derived using only the last
  two observations is shown as a star. See Sect. 4.2 for more details.}
%\label{fig:tprofile}
\end{figure}

\begin{figure}
\vskip 3truecm
\centerline{ \includegraphics[scale=0.7]{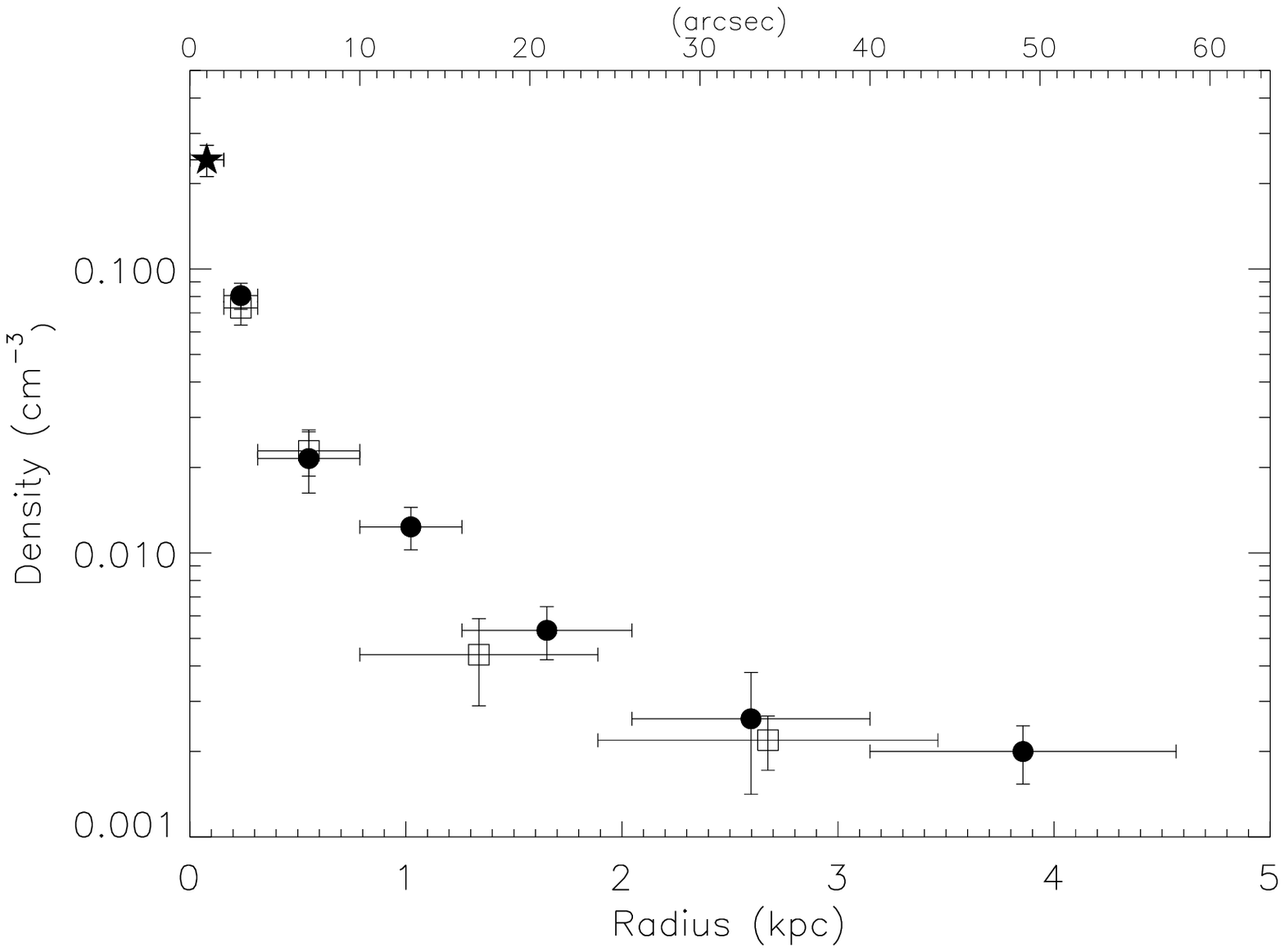} } 
\caption{Radial
  distribution of gas density derived from spectral deprojection along
  two directions (NE--SW, filled circles; and NW--SE, open boxes). The
  density for the inner 2\arcsec\/ derived using only the last two
  observations is shown as a star. See Sect. 4.2 for more details.}
%\label{fig:nprofile}
\end{figure}

\clearpage

\begin{figure}
\vskip -5.truecm
%\hskip -2. truecm
\includegraphics[height=11cm,width=11cm]{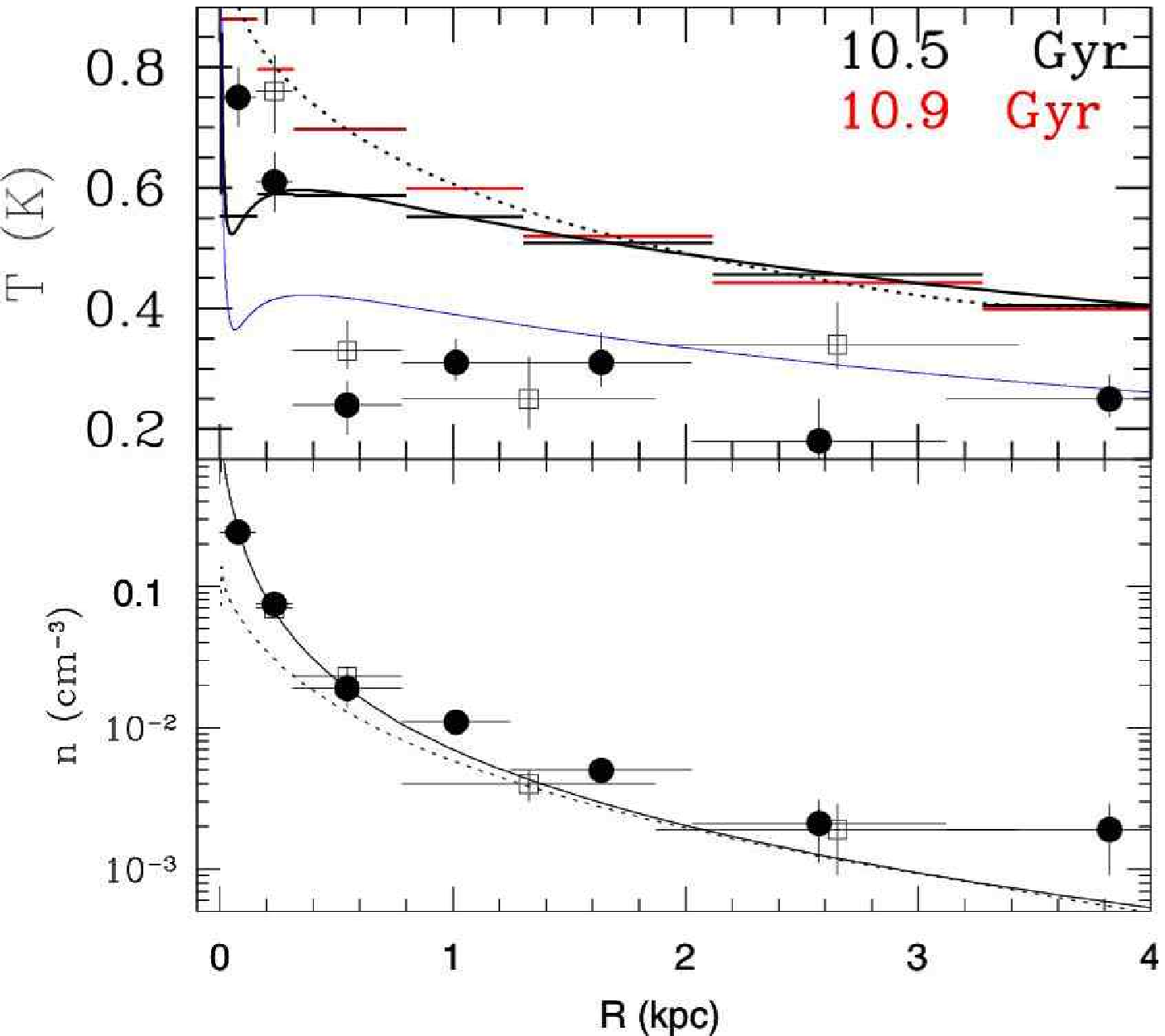}
%see /nfw35/paper
\caption{
The 3D radial temperature profile (upper panel) and density profile (lower
panel) for the observed values (full circles and open squares,
respectively for the NE-SW and NW-SE quadrants), and for the representative
model presented in the Appendix and discussed in Sect. 5.2.1, at two epochs: 10.5 Gyr (solid line) and 10.9 Gyr (dotted line). 
The model temperature at these epochs is also shown 
with bars, calculated by averaging the temperature with the 0.3--8 keV
emission over bins of a width as much as possible close to that used
for the observed data (compatibly with the spacing of the numerical grid).
The blue line shows the
heating due to stellar motions $T_{\sigma}(r)$; note the central spike due to the MBH.
See Sect.~\ref{nofeed} and the Appendix for more details. 
}
\label{hyd}
\end{figure}

\clearpage

\begin{table*}
\begin{center}
\caption{General properties of NGC4278}\label{tab1}
\medskip
%\scriptsize 
\begin{tabular}{ccccccccccc} 
\hline\hline 
  D    & Morph.& $B^T_0$&  $L_B/10^{10}$&Optical &  $M_{BH}$     &
  $\sigma_{e/8}$ & $R_e$\\
(Mpc)& type   &  (Bmag) &   $(L_{B,\odot}$)& class    &
($10^8M_{\odot}$) &     (km s$^{-1}$) & ($\arcsec$)   \\
  (1)  &  (2)    &   (3)       &   (4)                & (5)        &
  (6)          &          (7)   & (8)\\
\hline
16.1  & E1+  &  10.97    &       1.7           & L1.9     &  3.4
& 252 & 32\\
\hline \hline							
\end{tabular}							
\end{center}
Col. 1: distance from Tonry et al. (2001); cols. 2 and 3: morphological type and $B^T_0$ from RC3;
col. 4: $L_B$ from $B^T_0$ and the adopted distance;  col. 5: type of
optical nuclear emission (Ho et al. 1997); col. 6: $M_{BH}$ from the
$M_{BH}-\sigma$ relation (Tremaine et al. 2002); col. 7: stellar
velocity dispersion within a central aperture of $R_e/8$ (Kuntschner
et al. 2010); col. 8: effective radius (Cappellari et al. 2006).

\end{table*}

\begin{deluxetable}{lcccc}
\tabletypesize{\scriptsize}
%\rotate
\tablewidth{0pt}
\tablecaption{Summary of Nuclear Flux Variations\label{allflux}}
\tablehead{
\colhead{ObsId}   &
\colhead{Date}   &
\colhead{$\Gamma$}   &
\colhead{$F_{0.5-8\,{\rm keV}}$}   &
\colhead{$L_{0.5-8\,{\rm keV}}$}
}
\startdata
XMM-Newton & 2004 May 23 & 2.05~[2.03-2.07] & 33.8 [32.8-34.2]  & 10.70 [10.61-10.80] \\
Chandra 4741 & 2005 Feb 02 & 2.13~[2.00-2.28] & 18.1 [17.1-18.8]  &  5.74 [5.65-6.24] \\
Chandra 7077& 2006 Mar 16 & 2.26~[2.16-2.39] & 7.8 [7.5-7.9]   &  2.55 [2.48-2.79] \\
Chandra 7078& 2006 Jul 25 & 2.34~[2.22-2.47] & 16.0 [14.9-16.2] &  5.12 [4.99-5.74] \\
Chandra 7079& 2006 Oct 24 & 2.38~[2.28-2.50] & 14.1 [13.3-14.3]  & 4.67 [4.56-5.09] \\
Chandra 7081& 2007 Feb 20 & 2.12~[2.00-2.25] & 6.2 [5.9-6.4] & 1.99 [1.92-2.20] \\
Chandra 7080& 2007 Apr 20 & 2.02~[1.84-2.20] & 5.9 [5.5-6.1] & 1.86 [1.77-2.11] \\
Chandra 11269& 2010 Mar 15 & 2.31~[2.11,2.51] & 1.09 [0.98-1.15] & 0.36 [0.35-0.38]\\
Chandra 12124& 2010 Mar 20 & 2.31~[2.11,2.51] & 0.94 [0.65-1.01] & 0.29 [0.27-0.32]\\
\enddata

\tablecomments{$\Gamma$ is the photon index of the nuclear power law
  component; $F_{0.5-8\,{\rm keV}}$ is the observed flux in units
  of 10$^{-13}$ erg cm$^{-2}$ s$^{-1}$; $L_{0.5-8\,{\rm keV}}$ is the intrinsic
  luminosity in units of $10^{40}$ erg s$^{-1}$. Compiled
  from measurements in Younes et al. (2010) and this work. }
\end{deluxetable}

\begin{deluxetable}{cccccccc}
\tabletypesize{\scriptsize}
%\rotate
\tablewidth{0pt}
\tablecaption{X-ray Spectral Properties of the nucleus from the 2010 $Chandra$ observations\label{nuc_spec}}
\tablehead{
\colhead{$N_H$}   &
\colhead{$kT$ }   &
\colhead{$\Gamma$}   &
\colhead{$ L_{0.5-8{\rm keV}}$ }   &
\colhead{$ L_{APEC, 0.5-8{\rm keV}}$}   &
\colhead{$ L_{POW, 0.5-8{\rm keV}}$}   &
\colhead{$ L_{0.5-2{\rm keV}}$ }   &
\colhead{$ L_{2-10{\rm keV}}$}
\\
\colhead{(cm$^{-2}$)}   &
\colhead{(keV)}   &
\colhead{}   &
\colhead{(erg s$^{-1}$)}   &
\colhead{(erg s$^{-1}$)}   &
\colhead{(erg s$^{-1}$)}   &
\colhead{(erg s$^{-1}$)}   &
\colhead{(erg s$^{-1}$)}  }
\startdata
$4.18\pm3.13$ & $0.75\pm 0.05$ & $2.31\pm 0.20$ & 3.5E+39 & 1.0E+39 & 2.5E+39 & 2.5E+39  & 1.1E+39 \\
\enddata

\tablecomments{$\,$Luminosities are corrected for an absorption column of
  $N_H=4.18\times 10^{20}$ cm$^{-2}$; they refer to the total emission
  in the given band, when the spectral component is  not specified.}

\end{deluxetable}

\clearpage

\vskip -10truecm
\begin{deluxetable}{lccccc}
\tabletypesize{\scriptsize}
%\rotate
\tablewidth{0pt}
\tablecaption{Best-Fit Parameters for the Spectral Modeling with Deprojection\label{depro}}
\tablehead{
\colhead{Annuli}   &
\colhead{$N_H$}   &
\colhead{$kT$}   &
\colhead{$Norm_{kT}$}   &
\colhead{$\Gamma$}   &
\colhead{$Norm_{\Gamma}$}   
%\colhead{$L_{total}$}   &
%\colhead{$L_{APEC}$}  
%\colhead{$L_{LMXB}$}   &
%\colhead{$L_{CV+AB}$}
\\
\colhead{(arcsec)}   &
\colhead{($10^{20}$~cm$^{-2}$)}   &
\colhead{(keV)}   &
\colhead{}   &
\colhead{}   &
\colhead{}   
%\colhead{(erg s$^{-1}$)}   &
%\colhead{(erg s$^{-1}$)}   &
%\colhead{(erg s$^{-1}$)}   &
%\colhead{(erg s$^{-1}$)}
}
\startdata
 & & Nucleus (ObsID 11269,12124) & ($\chi^2/$dof=124.8/101) & & \\
\hline\\
0--2 & $4.18 \pm 3.13$ & $0.75\pm 0.05$ & 1.1E$-5\pm 1.3$E$-6$ & $2.31\pm 0.20$ &  2.1E$-5\pm 3.4$E$-6$ \\
\hline\\
 & & NE--SW (All ObsIDs) & ($\chi^2/$dof=692.8/616) & &  \\
\hline\\
2--4 &   1.76 &   $0.61(-0.05,+0.05)$ &   3.5E$-6\pm   3.6$E$-7$ & 1.8 & 4.5E$-6 \pm 4.1$E$-7$ \\
4--10 & 1.76 &   $0.24(-0.05,+0.04)$  &  2.5E$-6\pm   6.0$E$-7$ & 1.8 & 1.6E$-6 \pm 3.3$E$-7$ \\
10--16 & 1.76 & $0.31(-0.03,+0.04)$  &  4.0E$-6\pm   6.8$E$-7$ & 1.8 & 1.5E$-6 \pm 4.3$E$-7$ \\
16--26 & 1.76 & $0.31(-0.04,+0.05)$ &   3.6E$-6\pm   7.6$E$-7$ & 1.8 & 2.0E$-6 \pm 5.5$E$-7$ \\
26--40 & 1.76 & $0.18(-0.10,+0.07)$ &   3.0E$-6\pm   1.4$E$-6$ & 1.8  & 2.3E$-6 \pm 6.9$E$-7$ \\
40--58 & 1.76 & $0.25(-0.03,+0.04)$ &   5.1E$-6\pm   1.2$E$-6$ & 1.8 &  2.1E$-6 \pm 8.3$E$-7$ \\
\hline\\
 & & NW--SE (All ObsIDs) & ($\chi^2/$dof=409.5/350) & & \\
\hline\\
2--4   &  1.76 & $0.76(-0.07,+0.06)$ &   3.0E$-6\pm   3.9$E$-7$  & 1.8  &  3.8E$-6 \pm 4.6$E$-7$ \\
4--10 &  1.76 & $0.33(-0.03,+0.05)$ &    3.3$E-6\pm   6.0$E$-7$   & 1.8 &  3.2E$-6 \pm 4.0$E$-7$\\
10--24 & 1.76 & $0.25(-0.05,+0.07)$ &   2.1$E-6\pm   7.1$E$-7$   & 1.8  & 3.3E$-6 \pm 5.2$E$-7$ \\
24--44 & 1.76 & $0.34(-0.04,+0.07)$ &   3.3$E-6\pm   7.2$E$-7$   & 1.8  &  3.0E$-6 \pm 7.1$E$-7$ \\
\enddata
\tablecomments{Outside the nucleus, 
the fit required a very small absorption column ($N_H < 10^{20}$
  cm$^{-2}$), thus $N_H$ was fixed at the line-of-sight Galactic
  value towards NGC 4278 ($N_H=1.76\times 10^{20}$ cm$^{-2}$,  from
  the CXC tool COLDEN); also, the power-law $\Gamma$ was fixed at 1.8 to account for unresolved LMXBs.
 Solar abundance was assumed for the thermal component in all fits.
Only one $\chi^2$/dof is given for the series of annuli in each direction, because the spectral fitting was done simultaneously for the
annuli.}

\end{deluxetable}

\vskip 8truecm
\begin{deluxetable}{lccccc}
\tabletypesize{\scriptsize}
%\rotate
\tablewidth{0pt}
\tablecaption{Derived Physical Parameters for the Hot Gas Component \label{phys}}
\tablehead{
\colhead{Annuli}   &
\colhead{$n_e$}   &
\colhead{$p$}   &
\colhead{$E_{th}$}   &
\colhead{$L_{apec}$}   &
%\colhead{$\tau_{c}$}   &
\colhead{$M_{hot}$}   
%\colhead{$L_{total}$}   &
%\colhead{$L_{APEC}$}  
%\colhead{$L_{LMXB}$}   &
%\colhead{$L_{CV+AB}$}
\\
\colhead{(arcsec)}   &
\colhead{(cm$^{-3}$)}   &
\colhead{(dyne cm$^{-2}$)}   &
\colhead{(erg)}   &
\colhead{(erg s$^{-1}$)}   &
%\colhead{(10$^7$ yr)}   &
\colhead{(10$^5$ M$_{\odot}$)}   
%\colhead{(erg s$^{-1}$)}   &
%\colhead{(erg s$^{-1}$)}   &
%\colhead{(erg s$^{-1}$)}
}
\startdata
& & & Nucleus (ObsID 11269,12124) & & \\
\hline\\
0--2 & $ 0.242\pm0.030$&   5.8E-10 &  4.1E+53 &  1.0E+39  &     1.0\\
\hline\\
& & & NE--SW (All ObsIDs) & &  \\
\hline\\
2--4 &  $0.072\pm0.008$ &    1.4E-10 &   4.0E+53  &  1.6E+38  &      1.1 \\
4--10 & $0.019\pm0.005$ &    1.5E-11 &   6.9E+53 &   3.5E+38 &      5.0 \\
10--16 & $0.011\pm0.002$ &    1.1E-11 &   1.7E+54 &   1.7E+38  &       9.6\\
16--26 & $0.005\pm0.001$ &    4.7E-12 &   3.1E+54 &   1.2E+38 &    18.0 \\
26--40 & $0.002\pm0.001$ &    1.4E-12 &   3.2E+54 &   8.5E+37 &     30.3 \\
40--58 & $0.002\pm0.001$ &    1.4E-12 &   9.2E+54 &   8.4E+37 &    65.5 \\
\hline\\
& & & NW--SE (All ObsIDs) & &  \\
\hline\\
2--4 & $0.073\pm0.009$ &    1.8E-10  &   4.0E+53  &   9.7E+37  &    0.9 \\
4--10 & $0.023\pm0.004$ &    2.4E-11  &   9.1E+53  &   1.9E+38   &   4.8\\
10--24 & $ 0.004\pm0.001$ &    3.6E-12  &   1.8E+54  &   8.7E+37 &  12.6 \\
24--44 & $0.002\pm0.001$ &    2.3E-12  &   6.7E+54   &  6.8E+37  &  35.0\\
\enddata
\tablecomments{The electron number densities were derived from the emission measure for the thermal component.
Luminosities are reported for the 0.5--8 keV range, and are absorption
corrected ($N_H=1.76\times 10^{20}$ cm$^{-2}$).}
\end{deluxetable}

\clearpage
\section{Appendix}

The hot gas flow in NGC4278 has been investigated in
Sect.~\ref{nofeed} with spherically symmetric hydrodynamical
simulations following the evolution of the stellar mas losses from a
single burst passively evolving stellar population. This is a
reasonable model thanks to the regular optical appearance of NGC4278,
a roundish galaxy with a uniformely old stellar population (see the
Introduction). The galaxy model for the simulations, including a
central black hole, a stellar and a dark mass components, was built as
follows.  The stellar density was derived from the surface brightness profile 
in the $I$ band from $HST$ WFPC2 $F814W$ images and
large-field ground-based photometry (Cappellari et al. 2006).  This
profile extends from $152^{\prime\prime}$ down to $R\approx
0.05^{\prime\prime}$, within which the optical AGN dominates; within a
radius of $1^{\prime\prime}$ the profile flattens in a core; the
effective radius is $R_e=32^{\prime\prime}$ (equivalent to 2.496 kpc).
For the distance adopted here, the $I$-band luminosity is
$L_I=3.4\times 10^{10}L_{\odot,I}$, using the total observed magnitude
$I_T=8.83$, after correction for extinction (Cappellari et al. 2006).
\begin{figure}
% mass model of nfw35
\vskip -1truecm
%\hskip -0.5 truecm
\includegraphics[height=0.35\textheight,width=0.45\textwidth]{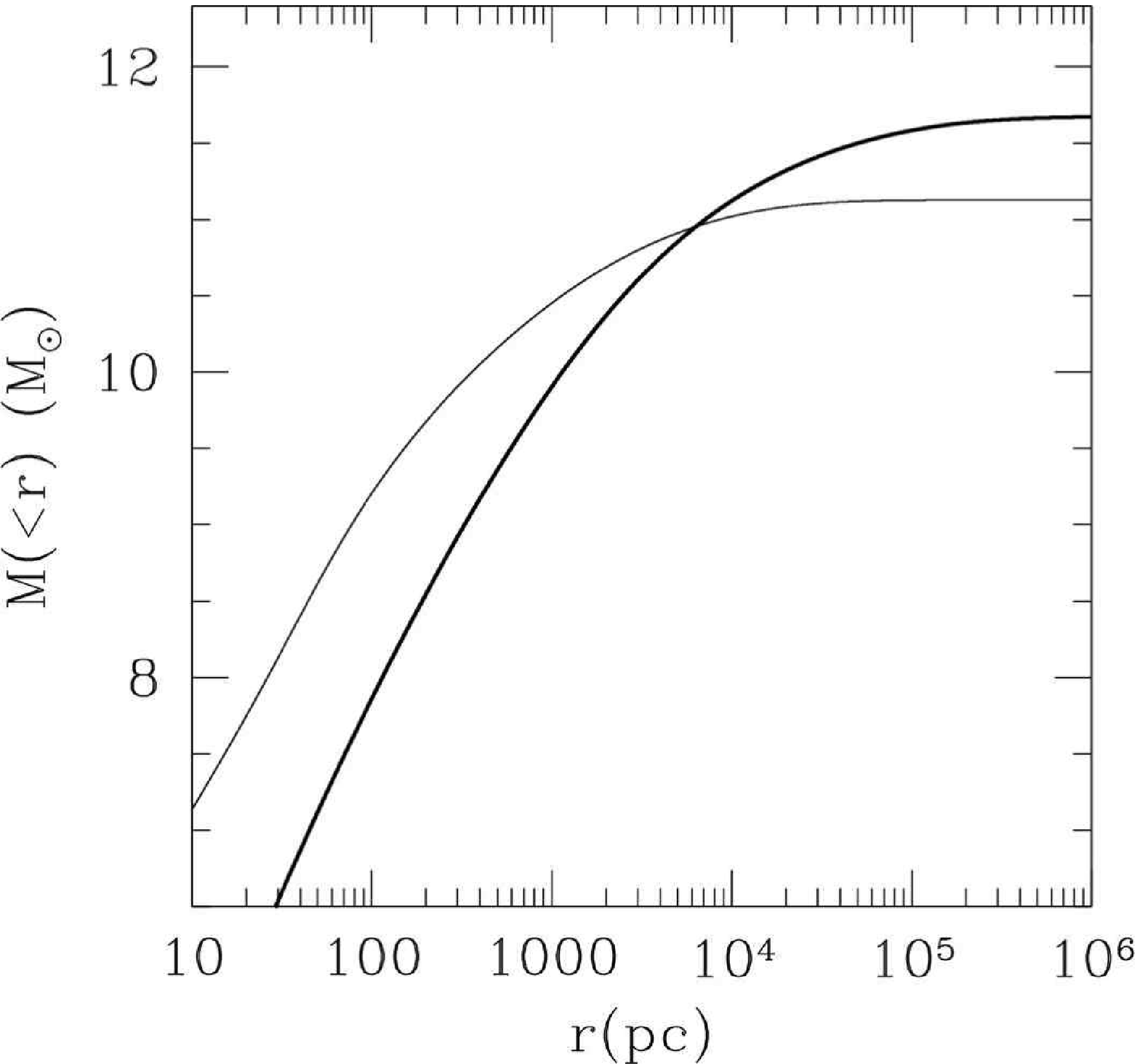}
%\vskip -0.5truecm
\vskip 1truecm
\includegraphics[height=0.35\textheight,width=0.45\textwidth]{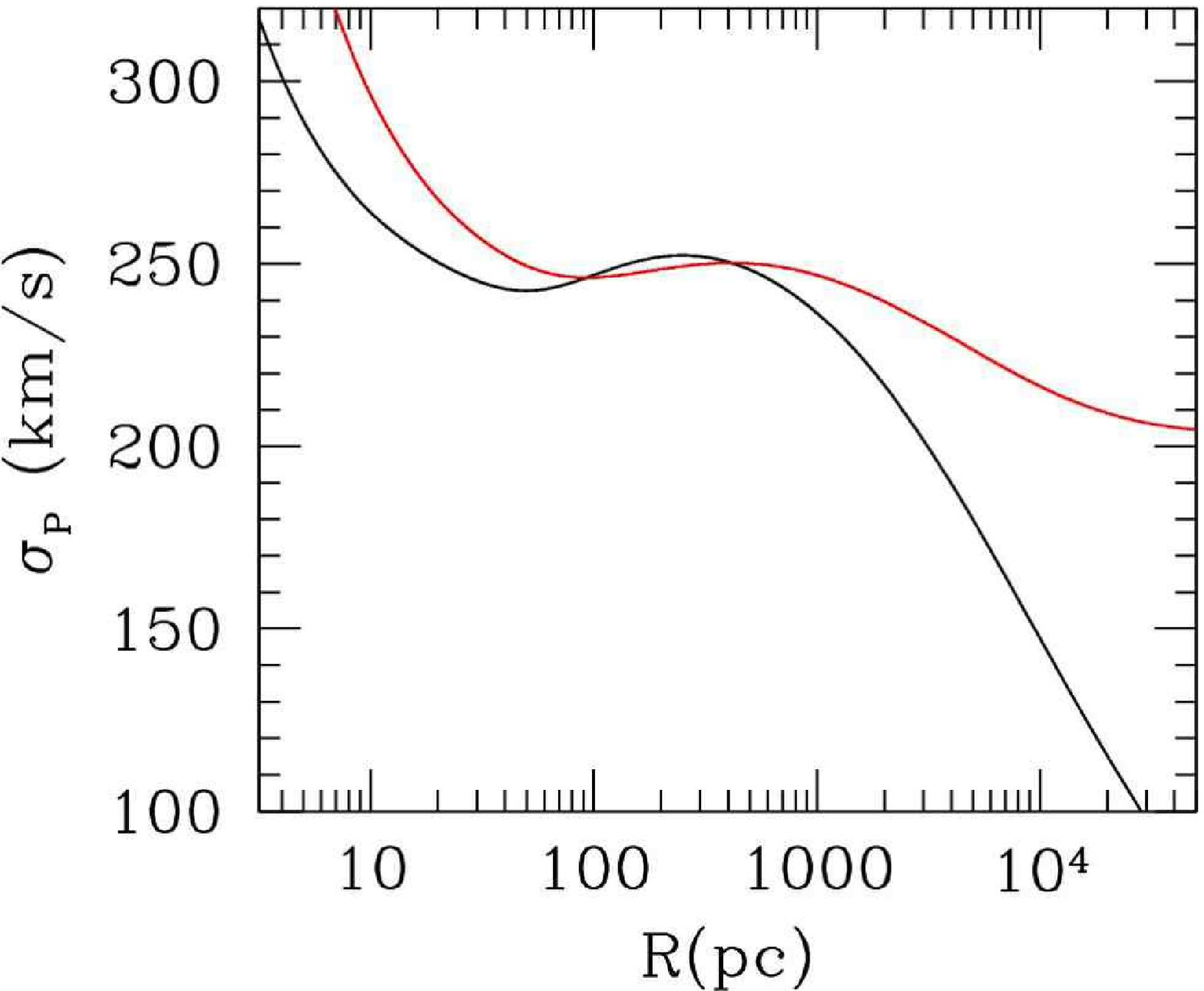}
%\vskip -1.5truecm
\caption{Top:  the mass profile for the adopted galaxy model (thicker line for the
dark mass, thinner one for the stellar mass). Bottom: the model
projected stellar velocity dispersion, in black, and the aperture
velocity dispersion, in red. Note how the red curve reaches the
aperture velocity dispersion of 252 km s$^{-1}$ at $R_e/8$ (and of
231$\pm 5$\% km s$^{-1}$ at $R_e$, as observed by Kuntschner et
al. 2010); note also the central increase produced by the MBH.}
\label{mass}
\end{figure}
By spherical deprojection of this surface brightness profile we
derived the 3D radial trend of the stellar density profile.  The
adopted central stellar velocity dispersion $\sigma_0$ was the
observed luminosity weighted velocity dispersion within a circular
aperture of radius $R_{e}/8$, that is $\sigma(<R_e$/8)=252 km s$^{-1}$
(Kuntschner et al. 2010; Tab.~\ref{tab1}). In the modeling of the
total mass distribution, we imposed this $\sigma_0$ value to the projected and
luminosity-weighted average of the stellar velocity dispersion within
a circle of radius $R_e/8$.

The dark halo was chosen to have a Navarro et al. (1997) profile [$\rho_h\propto
1/( r/r_h)(1+r/r_h)^2$, with $r_h$ the scale radius, and truncated at
large radii], and a total mass $M_h$.  The free parameters
$M_*,r_h,M_h$ are determined by imposing the $\sigma_0$ value, 
and the total mass-to-light ratio
within $R_e$ derived from 
models for NGC4278, $M/L_I=4.5-5.2$ (Cappellari et al. 2006), which
implies a dark-to-luminous mass ratio within $R_e$, (${\cal
 R}_e$), of $0.5-0.7$.  By solving numerically the Jeans equations for the
three mass components in the isotropic orbits case (e.g., Binney \&
Tremaine 1987; the stellar velocity dispersion of NGC4278 is nearly
isotropic, Cappellari et al. 2007), with a central MBH acting as a
point mass of $M_{BH}=3.4\times 10^8M_{\odot}$ (Tab.~\ref{tab1}), the resulting model
has a stellar mass-to-light ratio of $M_*/L_I=4.0$, ${\cal R}_e=0.69$,
${\cal R}=M_h/M_*=3.5$ (in good agreement with the value at 10$R_e$ of
$\sim 3$ from Bertola et al. 1993, Kronawitter et al. 2000).
Figure~\ref{mass} shows the main dynamical properties of the adopted mass
model.

\begin{figure}
%\vskip -1.truecm
%\hskip -2. truecm
\includegraphics[height=10cm,width=10cm]{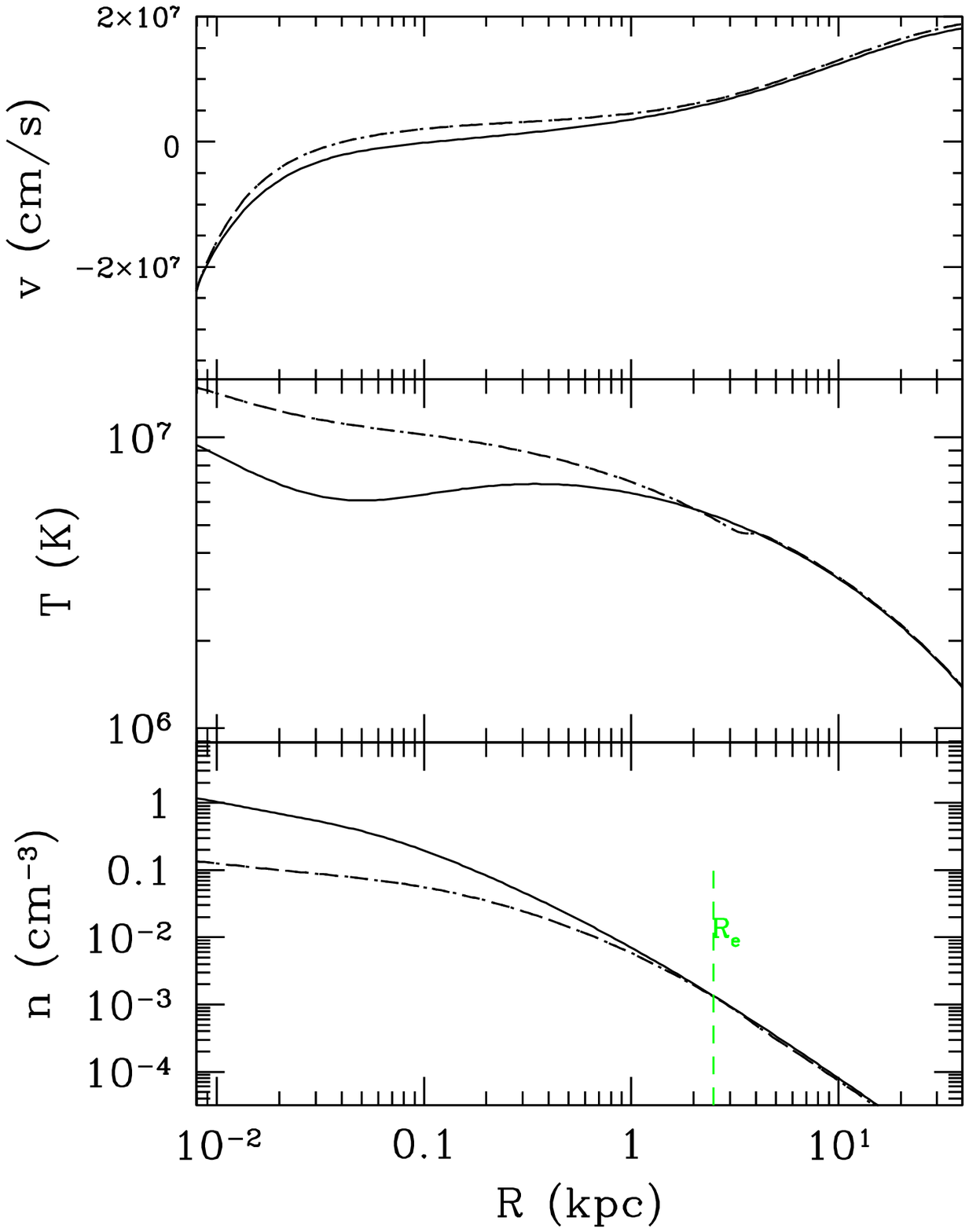}
%see /nfw35/paper
\caption{
Radial trend of the main hydrodynamical quantities of the gas flow for
the representative model described in the Appendix
and discussed in 
Sect.~\ref{nofeed}; displaied are the velocity (negative inward), top; the 
temperature, middle; the gas number density, 
bottom, for a galaxy age of 10.5 (solid) and 10.9 (dot-dashed) Gyr.
}
\label{hyd1}
\end{figure}

The time evolving input ingredients of the simulations are the rate of
stellar mass loss from the aging stellar population ($\dot M_*$), and
the rate of SNIa heating ($L_{SN}$).  The mass return rate prescribed
by the stellar evolution theory is used, as well as its decline as a
function of time, computed following the prescriptions of Maraston
(2005) for solar abundance and the Kroupa IMF (e.g., Pellegrini
2012). The present-epoch SNIa's rate for nearby early-type galaxies is
adopted, i.e., 0.16$h_{70}^2$ SNu (where 1 SNu = 1 SNIa per 100 yr
per $10^{10}L_{B,\odot}$, $h_{70}=H_{\circ}/70$; Cappellaro et
al. 1999; Li, W. et al. 2011).  The secular evolution of the rate is
taken as $\propto t^{-s}$, with $s=1.1$, as suggested by recent large
supernova surveys and modeling (e.g., Maoz et al. 2011).  Another
source of heating for the stellar mass losses is the thermalization of
the stellar random motions, that corresponds to a gas mass-weighted
temperature of $T_{\sigma}={1\over k} {\mu m_p \over M_*} \int 4 \pi
r^2 \rho_*(r) \sigma^2 (r) \,dr$, where $k$ is the Boltzmann
constant, $\mu m_p$ the mean particle mass,
$\sigma (r)$ is the one-dimensional velocity dispersion of the stars,
and $\rho_{gas}(r)\propto \rho_*(r)$ is assumed, where $\rho_*(r)$ is
the stellar density profile (see Pellegrini 2012 for more details).

The evolution of the galactic gas flow is obtained integrating the
time-dependent Eulerian equations of hydrodynamics with the above
described source terms, with a numerical code described in Pellegrini (2012),
with a sink of the hydrodynamical quantities at the
galactic center, and no feedback effects from the central MBH.  We
adopt a central grid spacing of 5 pc for a good sampling of the inner
regions, even within the Bondi radius, and a total of 240
logarithmically spaced gridpoints. The simulations begin at an age of
2 Gyr for the stellar population, when the galaxy has completed its
formation process and the galactic wind phase driven by type II supernovae
explosions has already started, and end at the age of NGC4278,
$10.7\pm 2.14$ (Terlevich \& Forbes 2002). Figure~\ref{hyd1}
shows the radial profiles of the main hydrodynamical quantities
(velocity, temperature and density) at the two epochs considered for
Fig. 8.

\end{document}